\documentclass[11pt,final]{article}
\pdfoutput=1
\usepackage{MRnotes}
\usepackage{hyperref}
\newcommand{\Ts}{{\sf T}}
\newcommand{\Js}{{\sf J}}

\newcommand{\uU}{\text{U}}

\newcommand{\beq}{\begin{equation}}
\newcommand{\eeq}{\end{equation}}
 
\newcommand{\bea}{\begin{eqnarray}}
\newcommand{\ea}{\end{eqnarray}}

\newcommand{\ii}{\mathrm{i}}
\renewcommand{\d}{\dd}
\renewcommand{\i}{\ii}

\title{Thermodynamics of the near-extremal Kerr spacetime}

 \author{Ilija Rakic${}^a$, Mukund Rangamani${}^a$, and Gustavo J. Turiaci${}^b$}
\affiliation{
	${}^a$ Center for Quantum Mathematics and Physics (QMAP)\\
	Department of Physics \& Astronomy, University of California, Davis, CA 95616 USA}
\affiliation{${}^b$ Physics Department, University of Washington, Seattle, WA USA.}
\emailAdd{irakic@physics.ucdavis.edu}
\emailAdd{mukund@physics.ucdavis.edu}
\emailAdd{turiaci@uw.edu}

\abstract{ 
We examine the thermodynamics of a near-extremal Kerr black hole, and demonstrate that the geometry behaves as an ordinary quantum system with a vanishingly small degeneracy at low temperatures. This is in contrast with the classical analysis, which instead predicts a  macroscopic entropy for the extremal Kerr black hole. Our results follow from a careful analysis of the gravitational path integral. Specifically, the low temperature canonical partition function behaves as $Z \sim \, T^\frac{3}{2}\, e^{S_0+ c \log S_0}$, with $S_0$ the classical degeneracy and $c$ a numerical coefficient we compute. This is in line with the general expectations for non-supersymmetric near-extremal black hole thermodynamics, as has been clarified in the recent past, although cases without spherical symmetry have not yet been fully analyzed until now. We also point out some curious features relating to the rotational zero modes of the near-extremal Kerr black hole background that affects the coefficient $c$. This raises a puzzle when considering similar black holes in string theory. Our results generalize to other rotating black holes, as we briefly exemplify. 
}

\begin{document}\maketitle
%~~~~~~~~~~~~~~~~~~~~~~~~~~~~~~~~~~~~~~~~~~~~~~~~~
\section{Introduction}\label{sec:intro}
%~~~~~~~~~~~~~~~~~~~~~~~~~~~~~~~~~~~~~~~~~~~~~~~~~

The semiclassical gravitational path integral approach to black hole thermodynamics pioneered by Gibbons and Hawking~\cite{Gibbons:1976ue}, while being a well-developed subject, continues to provide important new insights. A case in point, and the central focus of this paper, concerns near-extremal black holes, which provide an interesting laboratory for understanding quantum effects in semiclassical gravity~\cite{Iliesiu:2020qvm}. 

An extremal black hole with a degenerate Killing horizon, and thus a vanishing surface gravity, generically has  a non-vanishing horizon area. Naively, one has a non-vanishing entropy at zero temperature, viz., a system with macroscopic ground state degeneracy. In the absence of any symmetry, such a degeneracy should be lifted, see~\cite{Page:2000dk} for a nice discussion. Indeed, various ideas such as black hole pair production~\cite{Hawking:1994ii}, the attractor mechanism in string theory~\cite{Dabholkar:2006tb}, etc., have been advanced to argue for a vanishing degeneracy of non-supersymmetric extremal black holes. Furthermore, problems persist at low temperatures; the internal energy grows quadratically from its extremal value, $E-E_0 = T^2/M_\text{gap}$, naively implying the existence of a mass-gap of order $M_\text{gap}$ in the microscopic spectrum. Consequently, one anticipates that a near-extremal black hole at temperatures below this gap scale is unable to emit a single Hawking quantum in the canonical ensemble (fixed extensive charges)~\cite{Preskill:1991tb}. 

These decades-old puzzles have been clarified recently by a treatment of the gravitational path integral~\cite{Ghosh:2019rcj,Iliesiu:2020qvm,Heydeman:2020hhw}, with the desired upshot: one ends up removing the ground state degeneracy in the absence of an underlying symmetry principle. The key observation facilitating this is that the near-horizon region in the extremal limit includes a ${\rm AdS}_2$ part with enhanced $\mathrm{SL}(2,\mathbb{R})$ symmetry~\cite{Kunduri:2007vf}. This near-horizon region supports a set of large diffeomorphism (and gauge transformation) modes, which turn out to be  exact zero modes~\cite{Sen:2012kpz,Sen:2012cj} that are strongly coupled and, therefore, ought to be treated quantum mechanically. Fortunately, this task can be accomplished as described in~\cite{Stanford:2017thb} and reviewed in~\cite{Mertens:2022irh}, with the result being one-loop exact. Consequently, one has a clean prediction for the low temperature behaviour of the partition function; $Z(\beta) \sim T^\frac{3}{2}\, e^{S_0}$ where $S_0$ is the classical Bekenstein-Hawking entropy of the extremal solution.\footnote{
	As discussed in related works~\cite{Heydeman:2020hhw,Boruch:2022tno,Iliesiu:2022kny}, and also~\cite{Lin:2022zxd, Turiaci:2023wrh,Boruch:2023trc}, with sufficient supersymmetry the ground state degeneracy is protected. In this case, the system actually develops a mass-gap, with vanishing density of states between the BPS bound and the gap scale. This explains why microscopic counting of BPS black holes in string theory works.
}

This result, originally derived in~\cite{Ghosh:2019rcj} for the BTZ black hole (both from a bulk and also boundary CFT${}_2$ perspective) and in~\cite{Iliesiu:2020qvm} for charged Reissner-Nordstr\"om black holes using a dimensional reduction of the near-horizon region to JT gravity~\cite{Nayak:2018qej,Moitra:2018jqs}, can be understood quite generally in the framework of the gravitational path integral~\cite{Iliesiu:2022onk}. Since this discussion is universal, relying solely on the near-horizon  symmetry enhancement, it is interesting to examine the behaviour of the simplest extremal black hole, the Kerr black hole in vacuum Einstein gravity. There are two broad reasons for attempting this exercise:
\begin{itemize}[wide,left=0pt]
\item At a technical level, one can use the Kerr solution to test the general principles away from the spherically symmetric situation\footnote{A situation without spherical symmetry was considered in \cite{Boruch:2022tno}, where it is assumed based on symmetry arguments that the low energy behavior is described by a Schwarzian mode, even though it was not derived from first principles.}. While the dimensional reduction to JT gravity continues to apply at the classical level~\cite{Moitra:2019bub}, there are complications arising from the reduced symmetry. 
For one, the dimensional reduction to \AdS{2} is not quite natural, as this part of the geometry is warped and fibered  over a compact base space. For another, the low temperature corrections to the near-horizon region are complicated to describe in this language.  We therefore seek a direct computation of the quantum corrections, without relying directly on the connection to JT gravity. Our analysis is inspired by~\cite{Iliesiu:2022onk} and somewhat akin to the recent discussion of~\cite{Banerjee:2023quv} who undertake a similar exercise in the spherically symmetric context. 
 
\item The extremal Kerr entropy has been argued to be captured by a 2d chiral CFT, per the Kerr/CFT correspondence~\cite{Guica:2008mu}. Near-extremal thermodynamics was analyzed in~\cite{Castro:2009jf} and the developments in this area are reviewed in~\cite{Compere:2012jk}. The logic here was to leverage the near-horizon geometry of extreme Kerr spacetime~\cite{Bardeen:1999px}, which has a $\mathrm{SL}(2,\mathbb{R}) \times \mathrm{U}(1)$ isometry. By exploiting a particular set of large diffeomorphism modes, determined by suitable fall-off conditions, an asymptotic Virasoro algebra emerges from the diffeomorphisms of the azimuthal circle. The central charge is fixed by the black hole's angular momentum, and the extremal entropy is reproduced by the Cardy formula. The status of this approach deserves to be investigated in light of recent developments.
\end{itemize}

To contextualize our discussion, we recall that the first attempt to systematically understand the semiclassical quantum corrections around an extremal Kerr black hole was undertaken by Sen~\cite{Sen:2012cj}, who argued that the quantum corrected entropy of a Kerr black hole in pure Einstein-Hilbert gravity, with no additional matter degrees of freedom is given by
\begin{equation}\label{eq:senlog}
S = \frac{A}{4\,G_N} + \frac{64}{180} \,\log \frac{A}{4\,G_N} \,.
\end{equation}	
The logarithmic correction arises from the one-loop determinant of gravitons around the semiclassical black hole saddle.\footnote{
	One can analogously extract the leading correction for non-extremal black holes~\cite{Sen:2012dw}, obtaining instead $\frac{334}{180} \log A$. The two calculations do not have to agree on the coefficient of the $\log A$ term, as the near-extremal calculation is an asymptotic expansion at small temperatures (in the appropriate ensemble). A comparison with the non-extremal calculation would require a resummation of perturbative corrections in the temperature.}

\smallskip 

The corrections to~\eqref{eq:senlog} at low temperatures arise from two sources: a classical piece, which gives a linear temperature correction, and a quantum piece, which  arises from the zero mode contribution to the one-loop determinant. We are thus led to examining large diffeomorphisms in the near-horizon region of the extreme Kerr spacetime, the NHEK geometry of~\cite{Bardeen:1999px}. Armed with this, we then evaluate the determinant in the geometry with a small non-vanishing temperature. As explained in~\cite{Iliesiu:2022onk}, this provides an IR regulator, generating a non-vanishing action for the aforesaid zero modes, and enabling a clean estimate of their contribution to the determinant.

The zero modes in question are associated with large diffeomorphisms respecting the near-\AdS{2} asymptotics. Hitherto, two sets of zero modes were identified~\cite{Sen:2012cj}: one set of zero modes correspond to the Schwarzian dynamics of the \AdS{2} asymptopia, and the other set to the fluctuations of the black hole angular velocity. This can be motivated by dimensional reduction along the deformed two-sphere; the former modes correspond to gravitational fluctuations in \AdS{2}, while the latter is a zero mode of the ${\rm U}(1)$ Maxwell field, coming from the four-dimensional metric, obtained in the reduction. Working directly in the NHEK geometry, and examining the eigenspace of the spin-2 fluctuation operator (with a harmonic gauge fixing term), we are only able to identify the Schwarzian zero modes. The other mode, the rotational zero mode, whose form can be guessed from its physical effect, we find to\textbf{} be singular.\footnote{ The spectrum of fluctuations about the NHEK geometry was carefully analyzed in~\cite{Amsel:2009ev,Dias:2009ex}. However, these works focus on physical spin-2 fluctuations, which are best analyzed in terms of gauge-invariant degrees of freedom (the Teukolsky scalars adapted to the near-horizon). These gauge-invariant fields, however, do not capture the full set of large diffeomorphisms, which must be analyzed separately (for an illustration see Appendix E of~\cite{He:2022jnc}). } 
This poses an interesting puzzle, impacting the analysis of semi-classical corrections to rotating black hole thermodynamics.

Accounting for the Schwarzian modes alone as the only zero modes leads to the following result for the near-extremal entropy. In the canonical ensemble with fixed angular momentum we find
\begin{equation}\label{eq:sensch}
S = S_0+ \frac{154}{180} \,\log S_0  + 4\pi^2  \, \frac{T}{T_q} +  \frac{3}{2}\, \log \frac{T}{T_q} + \mathcal{O}({T^2})\,.
\end{equation}	
Here $S_0 = \frac{A_0}{4G_N}$  is the naive extremal entropy,\footnote{
	While it is helpful to view the result in terms of the entropy, note that the more natural presentation is directly in terms of the partition function, where it is clear that the result has to break down at very low temperatures owing to competition from other saddle point configurations. 
}
and $T_q$ is the emergent scale in the IR coming from the Schwarzian dynamics. The $\frac{3}{2} \, \log T$ term confirms a naive application of the general philosophy of~\cite{Iliesiu:2020qvm}. The discrepancy in the $\log S_0$ term relative to~\eqref{eq:senlog} is due to our only being able to identify the Schwarzian zero modes. We have chosen to present the result consistent with what we are able to conclusively establish. In the course of our analysis, we explain how the zero mode contribution affects the logarithmic corrections, and elucidate the sensitivity on the ensemble choice. Finally, the result~\eqref{eq:sensch} applies to four dimensional pure gravity, but it is easy to incorporate light matter fields in the evaluation of the quantum corrections; they only affect the coefficient of the $\log S_0$ term and not the temperature dependence.

\smallskip

The organization of the paper is as follows. We review the classical thermodynamics of the Kerr black hole in~\cref{sec:Kerr} following the Gibbons-Hawking prescription. In particular, we emphasize the choice of ensemble we use, as this will be crucial in ascertaining the low temperature corrections. Following this in~\cref{sec:LowT} we describe the geometry at low temperatures holding the  angular momentum fixed.  In~\cref{sec:QuKerr} we present our central result for the Kerr black hole low temperature thermodynamics accounting for the quantum effects from the near-horizon region. In~\cref{sec:genralizations} we generalize our results to other rotating black holes in four dimensions. Finally, we conclude in~\cref{sec:concl} with a broader discussion, focusing on the puzzles associated with the rotational zero modes, interpretation \'a la the  Kerr/CFT correspondence, etc., and outline some future directions for further investigation.

\bigskip

\emph{Note added:} While this work was in preparation,~\cite{Kapec:2023ruw} appeared on the arXiv. Their analysis overlaps with our results on the Schwarzian contribution to the near-extremal thermodynamics. The results of our analysis were announced in a recent talk at YITP, Kyoto~\cite{MYITP:2023sep}.

\bigskip 

\emph{Note added v2:} As indicated in~\cref{sec:concl}, we did not account for the instabilities of the Kerr black hole in our analysis. The near-extremal Kerr solution suffers from superradiance, which is reflected as an imaginary contribution to the effective action. From an estimate of the lifetime of the black hole, one concludes that the imaginary piece which sets the decay time is comparable to the one-loop contribution at temperatures of order  $T_q \sim J^{-3}$.  It would therefore appear that the near-extremal Kerr black hole radiates spin-2 gravitons and gets further from extremality before entering the regime where the quantum corrections kick in. This has an implication for the interpretation of our result as a statement concerning low-temperature thermodynamics. One can view our analysis as focusing on the absolute value of the one-loop effective action. We thank Juan Maldacena and Sameer Murthy for alerting us to this issue, and for helpful discussions.  

\bigskip

%~~~~~~~~~~~~~~~~~~~~~~~~~~~~~~~~~~~~~~~~~~~~~~~~~
\section{The Kerr black hole and its extremal limit}\label{sec:Kerr} 
%~~~~~~~~~~~~~~~~~~~~~~~~~~~~~~~~~~~~~~~~~~~~~~~~~

We begin with a brief review of some basic features of the classical thermodynamics of Kerr black holes, primarily to set our conventions. We will work with the simplest gravitational dynamics dictated by the Einstein-Hilbert action. In Euclidean signature, relevant for the evaluation of the gravitational path integral that computes the black hole free energy, the action is given by
\begin{equation}\label{eq:EHaction}
I_{_\text{EH}}
= 
	- \frac{1}{16 \pi G_N} \left[ \int_\mathcal{M} \dd[4] x\, \sqrt{g} R+ 2\, \int_{\partial \mathcal{M}}\, \dd[3]x\, \sqrt{\gamma} \,K \right] .
\end{equation}
In addition to the bulk Einstein-Hilbert action, we have included the Gibbons-Hawking-York (GHY) boundary term. This is  required when imposing Dirichlet boundary conditions on the metric (we will consider other boundary conditions, whence this boundary term will be appropriately  modified). One can add to this action contributions from minimally coupled matter fields. As we are going to study only geometric backgrounds, such matter fields will not affect the classical thermodynamics. 

%~~~~~~~~~~~~~~~~~~~~~~~~~~~~~~~~~~~~~~~~~~~~~~~
\subsection{The Kerr spacetime}\label{sec:kerr}
%~~~~~~~~~~~~~~~~~~~~~~~~~~~~~~~~~~~~~~~~~~~~~~

The general stationary axisymmetric black hole solution with a Killing horizon, is the Kerr black hole. The geometry has a non-vanishing ADM mass and angular momentum. In Boyer-Lindquist coordinates it is given by
\begin{equation}\label{eq:Kerr}
\dd s^2 
= 	
	- \frac{\rho^2 \Delta}{\Xi}\, \dd t^2 + \frac{\rho^2}{\Delta} \,\dd r^2 
	+ \rho^2 \,\dd \theta^2 + \frac{\Xi}{\rho^2}\, \sin^2 \theta\left(
	 \dd \varphi - \frac{2\,m\,a\,r}{\Xi}\, \dd t \right)^2 ,
\end{equation}
with 
\begin{equation}
\begin{split}
\rho^2 
&=
	 r^2 + a^2 \cos^2 \theta\,, \qquad 
\Delta
=
	 r^2 - 2 m r + a^2\,, \qquad 
\Xi 
= 
	(r^2 + a^2)^2 - a^2 \Delta \,\sin^2 \theta \,.
\end{split}
\end{equation}
The coordinates $(\theta,\varphi)$ parametrize a space that is topologically a sphere -- we  have  $\theta\in [0,\pi]$ while $ \varphi \sim \varphi + 2\pi$. The Euclidean geometry is obtained by Wick rotation $t \to - \ii  \tE$ and keeping $a\in \mathbb{R}$.\footnote{
	Thus, the Euclidean metric we have has a complex line element, which is indeed the correct saddle for the problem, cf.~\cite{Brown:1990fk}. } 
The metric is asymptotically flat since at large $r$ it approaches the Minkowski line element, with suitable subleading corrections.

To fully specify the boundary conditions, we need to decide a range for Euclidean time. Fixing temperature and angular velocity corresponds to the following identification
\begin{equation}\label{eq:tEphisim}
(\tE, \varphi) \sim (\tE+\beta,\varphi+ \ii \beta \Omega) \sim (\tE,\varphi + 2\pi).
\end{equation}
These boundary conditions fix the asymptotic metric, and the variational problem is well-defined with the GHY boundary term. As we will see later, fixing the angular momentum $J$ requires a different choice of boundary terms.

The relation between the thermodynamic data $\{T,\Omega\}$ and the parameters $\{m,a\}$ specifying the solution can be determined directly from the geometry. The radius of the outer horizon (i.e., the largest root of $\Delta$) is given by
\begin{equation}\label{eq:rp}
r_+ = m + \sqrt{m^2-a^2}.
\end{equation}
The Euclidean path integral requires a smooth geometry. At the horizon radius $r=r_+$, a combination of thermal and spatial circles becomes contractible. The resulting geometry has to be capped off smoothly, which fixes $r_+$ and $a$ in terms of the thermodynamic potentials.

Expanding the line element close to the horizon using $r= r_+ + \frac{(r_+^2 -a^2) \varepsilon^2}{4 r_+ (r_+^2 + a^2 \cos^2 \theta)}$, for small $\varepsilon$, we find the metric
\begin{equation}
\dd s^2 = 
	- \frac{(a^2-r_+^2)^2 \varepsilon^2 \dd t^2 }{4 r_+^2 (a^2+r_+^2)^2}+\dd \varepsilon^2 + (r_+^2 + a^2 \cos^2 \theta) \dd \theta^2 + \frac{(a^2 + r_+^2)^2\sin^2 \theta}{r_+^2 + a^2 \cos^2 \theta} (\dd \varphi - \frac{a}{r_+^2 + a^2} \dd t )^2 + \cdots, 
\end{equation}
where the ellipses indicate terms in the metric that are subleading in the near-horizon expansion. The smoothness at the poles of the sphere is guaranteed if we define $\widetilde{\varphi} = \varphi - \frac{a}{r_+^2 + a^2} \dd t$ and periodically identify $\widetilde{\varphi } \sim \widetilde{\varphi} + 2\pi$. Working at  $\theta$ and $\widetilde{\varphi}$ the thermal circle contracts at the horizon without a conical singularity when the Euclidean time is periodically identified, determining the temperature to be
\begin{equation}\label{eq:Temp}
T = \frac{r_+^2-a^2}{4\pi \,r_+ \,(a^2 + r_+^2)}=\frac{r_+ - m}{4\pi\, m \,r_+}\,.
\end{equation}
The coordinate identifications are $(\tE,\widetilde{\varphi}) \sim (\tE+\beta, \widetilde{\varphi})$. Reverting to the original coordinates this translates into $(\tE,\varphi) \sim (\tE + \beta, \varphi +\ii \frac{a}{r_+^2 + a^2} \beta)$, which determines the angular velocity
\begin{equation}\label{eq:OmKerr}
\Omega = \frac{a}{r_+^2 + a^2} = \frac{a}{2 \,m\, r_+}.
\end{equation}
Finally, the ADM mass and angular momentum evaluated for the Kerr metric~\eqref{eq:Kerr} are given by 
\begin{equation}\label{eq:MJKerr}
M = \frac{m}{G_N}, \qquad J=\frac{m \,a}{G_N}\,.
\end{equation}

\paragraph{Ensemble choices:} The action~\eqref{eq:EHaction} evaluated on the Kerr black hole with Dirichlet boundary conditions satisfies the quantum statistical relation 
\begin{equation}\label{eq:actionkerr}
I_{_\text{EH}} = - S + \beta \,M + \beta \,\Omega\, J\,,
\end{equation}
where the Bekenstein-Hawking entropy is given by the familiar formula
\begin{equation}\label{eq:SHBKerr}
S = \frac{A}{4 G_N} = \frac{2\pi\, m \,r_+}{G_N} \,.
\end{equation}
From this expression we can derive the thermodynamic interpretation of the quantities appearing in the r.h.s.~of~\eqref{eq:actionkerr}. This describes the physics in the grand canonical ensemble, holding $\beta$ and $\Omega$ fixed, and letting $r_+$ (or equivalently $m$) and $a$ to be adjusted accordingly. In this ensemble, the macroscopic charges $M$ and $J$, are derived quantities and can fluctuate. 

In the rest of the paper, we will work instead in an ensemble of fixed angular momentum. The importance of being explicit with the choice of ensemble can be made clear by reviewing the coordinate identifications. Recall that our the boundary conditions are specified by the identification~\eqref{eq:tEphisim}. This implies that $\ii \beta \Omega$ is only defined modulo $2\pi$ in a theory without fermions, and modulo $4\pi$ in a theory with fermions. The Kerr black hole solution, however, does not respect this symmetry. The resolution is that, when working at fixed angular velocity, there is actually an infinite family of solutions related by $\ii \beta \Omega \to \ii \beta \Omega +  2\pi \mathbb{Z}$ (for a theory without fermions) or $\ii \beta \Omega \to \ii \beta \Omega +  4\pi \mathbb{Z}$ (for a theory with fermions). These are genuinely different saddles with the same boundary conditions and should all be included in the evaluation of the partition function, as emphasized in~\cite{Chen:2023mbc}.

In an ensemble of fixed $J$, the canonical ensemble, this issue does not appear (there is a similar sum over saddles with the only effect of imposing the restriction that $2J\in \mathbb{Z}$). At the level of the classical analysis, the canonical ensemble requires a different boundary term in~\eqref{eq:EHaction} which we specify later. Its effect will essentially be to remove the $\beta\,\Omega\, J$ term on the r.h.s~of~\eqref{eq:actionkerr}.

%~~~~~~~~~~~~~~~~~~~~~~~~~~~~~~~~~~~~~~~~~~~~~~~~~
\subsection{A low temperature expansion}\label{sec:LowT}
%~~~~~~~~~~~~~~~~~~~~~~~~~~~~~~~~~~~~~~~~~~~~~~~~~

Let us examine the low temperature behavior of the Kerr black hole at fixed angular momentum $J$. We will take $J>0$ for concreteness. To simplify some expressions below, we introduce
\begin{equation}\label{eq:JTsdef}
\Js = \sqrt{G_N \,J}  \,, \qquad \Ts = 2\pi T.
\end{equation}
From~\eqref{eq:rp} and~\eqref{eq:Temp} in the extremal limit the mass and entropy are given by
\begin{equation}
M_0 = \frac{\Js}{G_N}\,, \qquad S_0 = \frac{2\pi \Js^2}{G_N}.
\end{equation}
Here and henceforth, we adhere to the convention that the extremal values are indicated by a subscript `0'. At non-vanishing temperature the mass and entropy receive corrections,
\begin{equation}\label{eq:SMne}
\begin{split}
M
&=
	\frac{\Js}{G_N} + \frac{\Js^3}{G_N} \, \Ts^2 + \cdots \,, \\ 
S
&=
	\frac{2\pi \,\Js^2}{G_N} + \frac{4\pi\, \Js^3 }{G_N}\, \Ts + \cdots\,.
\end{split}
\end{equation}
The ellipses indicate higher-order terms suppressed by additional factors of $\Js \Ts$. The temperature dependence of the leading correction to the extremal entropy and energy is quite universal. In particular, the fact that the mass approaches the extremal value quadratically leads precisely to the breakdown of the statistical mechanical interpretation of the black hole pointed out in~\cite{Preskill:1991tb}. More precisely, at low enough temperatures, the excitation energy of the black hole above extremality becomes comparable to the average energy of a Hawking quantum. This happens at a temperature scale 
\begin{equation}
T_q = \frac{G_N}{2 \,\Js^3}  \;\; \Longrightarrow \;\; \Ts_q = \frac{\pi \,G_N}{ \Js^3}.
\end{equation}
For temperatures $\Ts\gg\Ts_q$, the classical picture of the black hole is correct. At low temperatures $\Ts \sim \Ts_q$, quantum effects near the horizon become important, and the thermodynamic behavior has to be revisited. This will be the focus of the rest of the article.

\begin{figure}
\begin{center}
   \begin{tikzpicture}[scale=1.2, baseline={([yshift=0cm]current bounding box.center)}]
\draw[thick] (0,-2) ellipse (0.5 and 0.2);
\draw[thick] (-1,1) to [bend left=20] (-0.5,-0.5) -- (-0.5,-2) ;
\draw[thick] (1,1) to [bend right=20] (0.5,-0.5) -- (0.5,-2) ;
\draw[thick] (-3,0) -- (3,0) -- (3.5,2) -- (3.5-6,2) -- (-3,0);
\draw (2,2.25) node {\small Asymptotically 4d Flat};
\draw (1.2,-2) node {\small Horizon};
\draw[->,thick] (-1,1.3) -- (1,1.3);
\draw[->,thick] (1,1.3) -- (-1,1.3);
\draw (0,1.6) node {\small $\Js$};
\draw (1.3,-1) node {\small NHEK};
\fill[blue!50,nearly transparent] (-.74,.6) to [bend left=15] (-0.5,-0.5) -- (-0.5,-2) to [bend right=50] (.5,-2) -- (.5,-.5) to [bend left=15] (.74,.6) to [bend right =22] (-.74,.6);
\draw[thick,blue] (0,.6) ellipse (0.72 and 0.15);
\end{tikzpicture}
\end{center}
\caption{ A sketch of a spatial slice of the near-extremal Kerr black hole. The blue region is well approximated by the NHEK metric for low enough temperatures. Far from the horizon, the approximation breaks down and the metric transitions to the asymptotically flat extremal Kerr geometry.}
\label{fig:NHEK}
\end{figure}
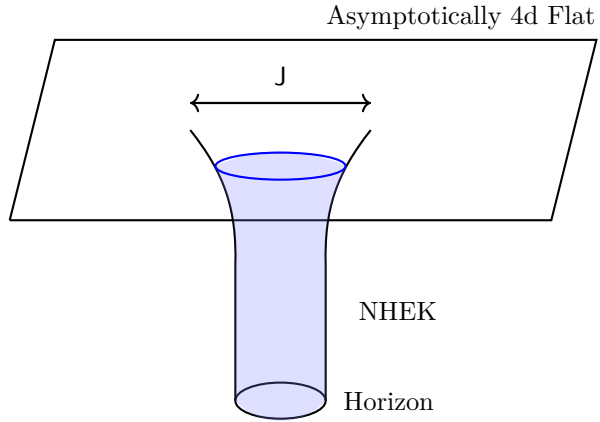

A different perspective on the results~\eqref{eq:SMne} can be obtained by looking at the Kerr geometry near-extremality. It is convenient to parametrize the coordinates by 
\begin{equation}
r= r_+ + 2\, \Js^2 \Ts (\cosh \rho -1)\,, \qquad \tau = \frac{\tE}{\Ts}\,.
\end{equation}
To leading order in $\Ts \to 0$ and keeping $\rho$ fixed, we obtain the NHEK geometry~\cite{Bardeen:1999px} 
\begin{equation}\label{eq:NHEKm1}
 \dd s_0^2 = (1+\cos ^2 \theta) \Js^2 \left[  \dd\rho^2 + \sinh^2\rho \,  \dd \tau^2 +  \dd \theta^2\right] + \frac{4 \,\Js^2 \,\sin^2 \theta}{1+\cos^2 \theta} ( \dd \varphi +\ii (\cosh\rho-1)\, \dd \tau)^2\,.
\end{equation}
Importantly, this is a fibration of \AdS{2} over $\vb{S}^2$. This is an exact solution of Einstein's equation~\cite{Bardeen:1999px} but gets corrections of order ${\sf J}{\sf T}$ coming from the full Kerr metric. 

From this perspective, the result~\eqref{eq:SMne} arises from studying the breaking of the symmetries \AdS{2}, deforming it by an irrelevant operator to connect to the asymptotically flat region. In fact, a quantum mechanical theory with an emergent conformal symmetry broken by finite temperature effects is described by the Schwarzian theory; it leads to the behavior~\eqref{eq:SMne}. At the classical level, the connection between the near-extremal Kerr black hole and the Schwarzian theory was analyzed in~\cite{Moitra:2019bub}.  We will justify the connection at the quantum level.   

It will be convenient to introduce another radial coordinate $y$ related to $\rho$ by $y= \cosh \rho$. In these coordinates, the NHEK metric becomes 
\begin{equation}\label{eq:NHEKm2}
 \dd s_0^2 
= (1+\cos ^2 \theta) \Js^2 \left[ (y^2-1) \dd \tau^2 + \frac{ \dd y^2}{y^2-1} +  \dd \theta^2\right] + \frac{4\, \Js^2\, \sin^2 \theta}{1+\cos^2 \theta} ( \dd \varphi +\ii (y-1) \dd \tau)^2\,.
\end{equation}
In the rest of the article we will go between~\eqref{eq:NHEKm1} and~\eqref{eq:NHEKm2} depending on which form is most convenient.

The NHEK geometry is obtained by keeping the radial coordinate $\rho$ (or equivalently $y$) fixed in the low temperature limit, and the conformal boundary of \AdS{2} is located at $\rho=\infty$. When analyzing quantum effects, it is important to work at a small but non-vanishing temperature. In this case, the approximations behind the NHEK geometry break down at $\rho_c \sim -\log(\Ts)$ before reaching the conformal boundary of \AdS{2}. When $\rho \sim \rho_c$ the near horizon geometry transitions to the extremal asymptotically flat Kerr metric. This behaviour is illustrated in~\cref{fig:NHEK}. 

%~~~~~~~~~~~~~~~~~~~~~~~~~~~~~~~~~~~~~~~~~~~~~~~~~
\subsection{A two-dimensional perspective on the low temperature limit}\label{sec:2dJT}
%~~~~~~~~~~~~~~~~~~~~~~~~~~~~~~~~~~~~~~~~~~~~~~~~~

For completeness, and to connect with the Schwarzian dynamics alluded to above, it is useful to recast the near-horizon solution in two-dimensional terms.  Focus first on the following metric ansatz
\begin{equation}\label{eq:ansatz4to2}
\dd s^2 = \frac{1+\cos^2\theta}{2\Phi}  \,g_{\mu\nu}\dd x^\mu \dd x^\nu +\frac{1+\cos^2\theta}{2}\Phi^2 \dd \theta^2 + \Phi^2  \frac{2\sin^2\theta}{1+\cos^2\theta}  (\dd \varphi - \ii A)^2.
\end{equation}
Here $x^\mu = (\tau,\rho)$ and $g_{\mu\nu}$ is a two-dimensional metric along these directions alone. The `dilaton' $\Phi(\tau,\rho)$, the one-form $A=A_\mu \dd x^\mu$ specifying a $\uU(1)$ connection, and the metric $g_{\mu\nu}$ depend only on time and radial coordinates. The dependence on the polar coordinate $\theta$ is fixed by the ansatz, and $\partial_\varphi$ is a Killing field. 

The four dimensional Einstein-Hilbert action can be reduced to a two-dimensional dynamical system along $\tau$ and $y$. Focus, in particular, on the fields, appearing in~\eqref{eq:ansatz4to2}.\footnote{ 
	Any mode that does not fall in this category, can be viewed as a matter field in two  dimensions, as long as they respect the polar angle dependence specified by the ansatz. This issue will be important in the sequel.} 
The two-dimensional metric, dilaton and gauge field have their dynamics specified by
\begin{equation}\label{eq:ehtodg}
I_{_\text{EH}} =	
	 -\frac{1}{4}\, \int_{\mathcal{M}_2} \dd^2 x \,\sqrt{g}\left[ \Phi^2\, R + \frac{1}{\Phi} +\frac{1}{2} \Phi^5 \,F^2\right]	
	-\frac{1}{2} \int_{\partial\mathcal{M}_2} \, \sqrt{h}\, \Phi^2 K.
\end{equation}%

This is a two-dimensional dilaton-gravity theory with a particular dilaton potential. The boundary term corresponds to fixing the boundary two-dimensional metric and $\uU(1)$ gauge connection.

It is easy to verify that the NHEK line element~\eqref{eq:NHEKm2} is a solution of~\eqref{eq:ehtodg}, with the two-dimensional metric reducing to the constant negative curvature metric on \AdS{2}, viz.,
\begin{equation}
g^{(0)}_{\mu\nu} \dd x^\mu \dd x^\nu =\Phi_0^{3/2} (\dd \rho^2 + \sinh^2 \rho \, \dd \tau^2 )\,.
\end{equation}
The dilaton is constant $\Phi_0 = 2 \Js^2$ and the $\uU(1)$ connection is a left-invariant one-form on \AdS{2}
\begin{equation}\label{eq:2dA}
A^{(0)} = - (\cosh \rho -1) \,\dd \tau.
\end{equation}

The two-dimensional perspective is useful to define more precisely what we mean by a fixed angular momentum ensemble, at least in the NHEK region. The angular momentum is proportional to the conserved charge that couples to the $\uU(1)$ gauge field $A$. Fixing the angular momentum is then equivalent to fixing the charge of the gauge field at the boundary. To make the variational problem of the action~\eqref{eq:ehtodg} well-defined, it is necessary to add a new boundary term given by   
\begin{equation}
I_{_\text{EH}} \to I_{_\text{EH}} +\frac{1}{2} \int_{\partial\mathcal{M}_2} \,\sqrt{h}\, n^\mu\, \Phi^5\, F_{\mu\nu} A^\nu\,,
\end{equation}
where $n^\mu$ is a normal to the two-dimensional boundary and $h$ the induced metric. One can verify that in the NHEK solution this term is precisely $\beta \Omega J $. Upon integrating over the boundary holonomy it implements the Legendre transform required to go to fixed charge ensemble. This boundary term can be directly written in four dimensional language using the results from~\cite{Brown:1992bq}.

%~~~~~~~~~~~~~~~~~~~~~~~~~~~~~~~~~~~~~~~~~~~~~~~~~
\section{Quantum effects at low temperatures}\label{sec:QuKerr} 
%~~~~~~~~~~~~~~~~~~~~~~~~~~~~~~~~~~~~~~~~~~~~~~~~~

We now turn to the evaluation of quantum corrections to the black hole entropy of the (near) extremal Kerr black hole. In order to do so, we will follow Sen and focus on the path integral in the near horizon region (see~\cite{Sen:2012kpz} for a nice presentation for the case of charged black holes). We shall begin our analysis by considering the zero temperature $\Ts=0$ black hole, and subsequently incorporate temperature corrections. 

%~~~~~~~~~~~~~~~~~~~~~~~~~~~~~~~~~~~~~~~~~~~~~~~~~
\subsection{Quantum effects in extremal Kerr}
%~~~~~~~~~~~~~~~~~~~~~~~~~~~~~~~~~~~~~~~~~~~~~~~~~

Let us focus on the near-horizon NHEK geometry~\eqref{eq:NHEKm2} of the extremal Kerr spacetime. The quantum effects we want to compute are captured by the one-loop determinant of the graviton and matter fields around this background geometry.

We begin by explaining the general strategy to compute the quantum corrections around this (or any) geometry. Let $x=(\tau,y,\theta,\varphi)$ denote the four dimensional coordinates. The one-loop determinant may  be evaluated by expanding the metric fluctuations (or those of other matter fields) in a complete orthonormal basis of normalizable modes $f^{(i)}_n(x)$. Here $i$ runs over a complete set of degrees of freedom at each  spacetime point. 

Consider then the heat kernel $K^{ij}(x,x';s)$ defined using the eigenspectrum of the kinetic operator 
\begin{equation}
K^{ij}(x,x';s) = \sum_n \, e^{-\lambda_n s} \, f_n^{(i)}(x) \, \bar{f}_n^{(j)}(x')\,.
\end{equation}
The basis has been suitably chosen to diagonalize the kinetic operator for the fluctuations (derived from the Gaussian approximation of the action) with  eigenvalues $\lambda_n$. For bookkeeping purposes it is convenient to include in this sum potential zero modes with $\lambda_n=0$, and correct for their effect at the end of the calculation. The contribution to the one-loop determinant of all fluctuations is then given by
\begin{equation}\label{eq:ZoneloopK}
\log Z_{\rm 1-loop} = \frac{1}{2} \, \int_{\Lambda_{\rm UV}}^\infty \, \frac{\d s}{s} \int \d^4 x\,  \sqrt{g} \, K(x,x;s)\,.
\end{equation}
The integral over $s$ requires a UV cut-off $\Lambda_{\rm UV}$. The heat kernel admits a Laurent series expansion in powers of $\sqrt{s}$ around $s=0$. In even dimensions, only even powers of $\sqrt{s}$ appear, while in odd dimensions only odd powers of $\sqrt{s}$ appear. Let $K_0(x)$ the constant term in this expansion, following~\cite{Sen:2012cj} -- it only appears in even dimensions, which is the case of interest for us ($d=4$). $K_0(x)$ is an important quantity, since it leads to a term in the one-loop determinant that grows logarithmically with the area of the black hole, or equivalently in our case with ${\sf J}$. To see this we can replace the contribution from $K_0$ in equation~\eqref{eq:ZoneloopK} and notice that the $s$ integration region $\Lambda_{\rm UV} \ll s \ll {\sf J}^2$ leads to $\log ({\sf J}^2/\Lambda_{\rm UV}) = \log (S_0) + \order{1}$, using that the UV cut-off is of order $G_N$. Therefore, in the macroscopic limit, the logarithmic correction is the leading quantum effect to the entropy. 

For pure gravitational background such as the NHEK geometry, it is easy to compute $K_0(x)$ even in the presence of matter fields. Since the matter fields are not turned on the background~\eqref{eq:NHEKm2}, there is no kinetic mixing to quadratic order between the matter fields and the graviton. In the presence of $n_S$ minimally coupled massless scalar fields, $n_V$ minimally coupled massless vector fields, $n_F$  minimally coupled massless Dirac fields, and a single spin-2 field, $K_0$ is given by~\cite{Duff:1977ay,Christensen:1978md,Christensen:1979iy,Duff:1980qv, Vassilevich:2003xt}
\begin{equation}
K_0 (x) = - \frac{1}{90 \pi^2}\, (n_S + 62 \, n_V + 11\,  n_F) E - \frac{1}{30\pi^2}\,  (n_S + 12\, n_V + 6\,  n_F + \frac{424}{3}) \,I\,,
\end{equation}
where 
\bea
E &=& \frac{1}{64} (R_{\mu\nu\rho\sigma}R^{\mu\nu\rho\sigma} - 4 R_{\mu\nu}R^{\mu\nu} + R^2),\\
I &=& -\frac{1}{64} (R_{\mu\nu\rho\sigma}R^{\mu\nu\rho\sigma} - 2 R_{\mu\nu}R^{\mu\nu} + \frac{1}{3}R^2).
\ea
The invariants appearing in $K_0$ are straightforward to compute, as only the Kretschmann scalar $R_{\mu\nu\rho \sigma} R^{\mu\nu\rho \sigma}$ is non-vanishing on solutions of vacuum Einstein's equations. 
Integrating $K_0(x)$ over the spacetime geometry, we are left with the  $s$ integral, which produces a logarithmic term in ${\sf J}$, as explained in~\cite{Sen:2012cj} and recalled in the previous paragraph. The final answer is
\begin{equation}\label{eq:logZp}
\log Z'_{\rm 1-loop} = \frac{1}{180}  \left( 2\, n_S - 26 \,n_V + 7 \,n_F  + 424\right) \log S_0 +\cdots\,,
\end{equation}
where the prime indicates that in this result we haven't yet accounted for the zero modes. The ellipses correspond to terms that are either order one or decay in the large $S_0$ limit and are therefore sub-leading. In particular, note that the contribution from gravitons gives a factor of $\frac{424}{180} \log S_0$. We have to correct this by accounting for the graviton zero modes of the background.

%~~~~~~~~~~~~~~~~~~~~~~~~~~~~~~~~~~~~~~~~~~~~~~~~~
\subsection{The zero modes of extremal Kerr}
%~~~~~~~~~~~~~~~~~~~~~~~~~~~~~~~~~~~~~~~~~~~~~~~~~

Our main goal is to revisit the zero mode contributions. In order to do so, we should first understand how they arise, and  then  characterize them explicitly. On general grounds in the extremal Kerr metric, one expects two sets of such modes~\cite{Sen:2012cj}, which can be distinguished by their transformation under  $\mathrm{SL}(2,\mathbb{R})$ isometry, cf.,~\cite{Camporesi:1994ga}. The first of these are usually referred to as the tensor zero modes in the literature.  They are associated with the asymptotic symmetry of \AdS{2}. We will therefore refer to them as the \emph{Schwarzian modes}. The second set of modes are associated with the ${\rm U}(1)$ gauge field obtained under the dimensional reduction to \AdS{2}. They are usually referred to as the vector zero modes in the literature, but since they respect the axisymmetry of the Kerr solution, we will refer to them as the \emph{rotational modes}. It is worth emphasizing that both these modes are associated with metric perturbations in the four dimensional NHEK geometry.

In order to determine the modes in question, we need to construct the quadratic fluctuation operator for the Einstein-Hilbert theory. We will do so following~\cite{Christensen:1979iy}, by explicitly including a gauge fixing term, and expanding the resulting action to Gaussian order in perturbations about the NHEK geometry. We pick the harmonic gauge fixing term and let 
\begin{equation}\label{eq:gaugefix}
I_{_\text{grav}} = I_{_\text{EH}} + I_\text{gf}  \,, \qquad 
I_\text{gf} = \frac{1}{32\pi G_N}\, \int \d^4x\, \sqrt{g} \left(\nabla_\mu h\indices{^\mu_\nu} - \frac{1}{2}\, \nabla_\nu h\right)\left(\nabla_\rho h\indices{^{\rho\nu}} - \frac{1}{2}\, \nabla^\nu h\right),
\end{equation}	
where $I_{_\text{EH}}$ was given earlier in~\eqref{eq:EHaction}. Expanding this to quadratic order, we obtain
\begin{equation}
I_{_\text{grav}}^{(2)} = \frac{1}{16\pi G_N}\, \int \d^4x\, \sqrt{g} \, \frac{1}{4}\,  \widetilde{h}^{\mu\nu} \Delta_L\, h_{\mu\nu}\,,
\end{equation}
with $\widetilde{h}_{\mu\nu}  = h_{\mu\nu}- \frac{1}{2}\, g_{\mu\nu}\, h$.  The operator $\Delta_L$ is the spin-2 Lichnerowicz operator, which maps symmetric rank-2 tensors to symmetric rank-2 tensors via\footnote{ If we have a non-vanishing cosmological constant, then $\widetilde{h}^{\mu\nu}(\Delta_L^\Lambda h_{\mu\nu}) = \widetilde{h}^{\mu\nu}(\Delta_Lh_{\mu\nu}) - 2\, \Lambda \widetilde{h}^{\mu\nu}h_{\mu\nu}$.	}
\begin{equation}\label{eq:DeltaLgen}
\begin{split}
\Delta_L h_{\mu\nu} = -\nabla_\rho \nabla^\rho h_{\mu\nu} + 2 \, R_{\mu\rho}\,h\indices{^\rho_\nu} -2\, R_{\mu\rho\nu\sigma}\, h^{\rho \sigma}  
- 4\left(R_{\nu\sigma}-\frac{1}{4}\, g_{\nu\sigma}\, R\right) h\indices{^\sigma_\mu} \,.
\end{split}
\end{equation}	
Since the Kerr spacetime is Ricci-flat, we can simplify the operator to 
\begin{equation}\label{eq:DeltaLRflat}
\Delta_L = -\nabla_\rho \nabla^\rho h_{\mu\nu}  -2\, R_{\mu\rho\nu\sigma}\, h^{\rho \sigma} \,.
\end{equation}

We can now give a mathematical characterization of the problem. The construction of the heat kernel proceeds by finding a complete set of eigenmodes for this Lichnerowicz operator. However, as detailed above, we circumvent the explicit analysis of the eigenspectrum by using the fact that the leading large charge correction to the entropy can be extracted from the conformal anomaly. Since we have included the zero modes in that analysis, we need to isolate their contribution. The zero modes in question are metric perturbations that lie in the kernel of $\Delta_L$. We will also require that such perturbations be normalizable with respect to the ultralocal measure ${\rm D}h$, which is defined by the property
\begin{equation}
\int {\rm D}h \, e^{-\int\, \d^4x\, \sqrt{g}\, \widetilde{h}_{\mu\nu}\, h^{\mu\nu} } =1\,.
\end{equation}

The Faddeev-Popov ghosts that come with the gauge fixing action $I_\text{gf}$ are implicitly included in the derivation of~\eqref{eq:logZp}. As noted in the literature, cf.,~\cite{Sen:2012kpz}, there are no zero modes arising from the ghost sector. For our discussion, which seeks to revisit the zero mode contribution to the one-loop determinant, we do not have to worry about the ghost modes. For further details the reader should consult the aforementioned earlier literature. 

%~~~~~~~~~~~~~~~~~~~~~~~~~~~~~~~~~~~~~~~~~~~~~~~
\subsubsection{The Schwarzian zero modes}\label{sec:tensorz}
%~~~~~~~~~~~~~~~~~~~~~~~~~~~~~~~~~~~~~~~~~~~~~~

First, let us consider the Schwarzian zero modes, whose presence owes to the \AdS{2} asymptotics.  The fastest way to find them is by direct computation of a metric perturbation that is annihilated by the Lichnerowicz operator. Consider a linear fluctuation of the NHEK line element by  $\dd s_\text{Sch}^2$, viz., 
\begin{equation}
\d s^2 = \d s_0^2 + \d s_\text{Sch}^2 \,.
\end{equation}
By plugging in an ansatz for the fluctuation to  have no components along the angular directions, and the fields to be functions only of the \AdS{2} coordinates, one can directly solve for the modes in question. One finds that the Schwarzian modes takes the following form:
\begin{equation}
\d s_\text{Sch}^2 
= \Js^2 \, (1+\cos^2\theta) \left[ H_{\tau \tau}(\tau,y) \, \d \tau^2 + 2\, H_{\tau y}(\tau, y) \, \d \tau \,\d y + H_{yy}(\tau, y)\,  \d y^2 \right] , 
\end{equation}
where the functions $H_{\mu\nu}(\tau,y)$ are given explicitly by
\begin{equation}\label{eq:zeromodes}
\begin{split}
H_{\tau\tau}(\tau,y) 
&= 
	- \left( \frac{y-1}{y+1}\right)^{\frac{n}{2}} \left( \delta_+ \, e^{\i n \tau} + \delta_-\,  e^{- \i n \tau}\right) , \\
H_{\tau y}(\tau,y) 
&=
	\i \frac{1}{y^2-1} \left( \frac{y-1}{y+1}\right)^{\frac{n}{2}} \left( \delta_+ \, e^{\i n \tau} - \delta_-\, e^{- \i n \tau}\right) ,\\
H_{yy}(\tau,y) 
&= 
	\frac{1}{(y^2-1)^2} \left( \frac{y-1}{y+1}\right)^{\frac{n}{2}}  \left( \delta_+ \,e^{\i n \tau} + \delta_- \,e^{- \i n \tau}\right).
\end{split}
\end{equation}

Within the linearized ansatz the parameters $\delta_\pm$ are in principle arbitrary. To ensure reality of the metric fluctuation, we impose $\delta_+ \propto \xi_n$ and $\delta_-\propto \overline{\xi}_{n}= \xi_{-n}$. The normalization of the modes can be determined by demanding orthonormality of the eigenfunctions of the spin-2 Laplace operator, which we shall specify precisely in the following section, cf.,~\eqref{eq:delpmvals}. We will also shortly argue that $n \in \mathbb{N}/ \{0,1\} $.  

A simpler way to determine these modes is to realize that arise from large diffeomorphisms in \AdS{2}. They correspond precisely to the Schwarzian boundary graviton of~\cite{Jensen:2016pah, Maldacena:2016upp, Engelsoy:2016xyb}, see~\cite{Mertens:2022irh} for a review. Picking an ansatz for the vector field generating the diffeomorphism that exploits the underlying  symmetry of the NHEK geometry,  say (for simplicity focusing on the positive frequency modes)
\begin{equation}\label{eq:zetaT}
\zeta_T = e^{\i n \tau} \left(f_1(y)\, \pdv{y} + f_2(y) \pdv{t} + f_3(y) \pdv{\varphi} \right), 
\end{equation}
one can again directly solve for the resulting metric perturbation to lie in the $\ker(\Delta_L)$. This fixes
\begin{equation}
f_2(y) =  \frac{\i\, f_1'(y)}{n}\,, \qquad f_3(y) =  \frac{(y-1)\,f_1'(y)-f_1(y)}{n}\,, 
\end{equation}
and leads to a ODE for $f_1(y)$, which is solved straightforwardly and determines 
\begin{equation}\label{eq:f1}
f_1(y) =\left(\frac{y-1}{y+1} \right)^\frac{n}{2}\,(n+y) \,\frac{\delta_+}{2\,(n^2-1)} \,.
\end{equation}
This solution is actually the unique smooth vector zero mode generated by a diffeomorphism of the form~\eqref{eq:zetaT}. 

\smallskip

Since this is an important fact (in particular in regard to the discussion in the next section) we explain this in some detail before moving on. First, one can evaluate $$\Delta_L h_{\theta\theta} |_{\theta=0} - \Delta_L h_{\theta\theta} |_{\theta= \frac{\pi}{2} }= - 6 (f_1'(y) + \i n f_2(y))=0,$$ imposing the first relation above. To find the solution for $f_3(y)$, we can write $f_3(y) =  \frac{(y-1)\,f_1'(y)-f_1(y)}{n} + \delta f_3(y)$ and evaluate then the following component of the Lichnerowicz operator $$\partial_\theta (\Delta_L h_{\theta y} )|_{\theta=0}+\frac{1}{2}\partial_\theta (\Delta_L h_{\theta y} )|_{\theta=\frac{\pi}{2}} = 14 n e^{\i n \tau} (y^2-1)^{-1} \delta f_3(y),$$ which automatically implies that $\delta f_3(y)=0$. Inserting this in any of the remaining non-trivial components of $\Delta_L h$ produces a second order differential equation for $f_1(y)$. One solution was quoted in~\eqref{eq:f1} while the other (obtained by replacing $n\to -n$) is singular at the horizon $y=1$ for $n>0$. Therefore, the Schwarzian zero mode is the only one of the form~\eqref{eq:zetaT}.

\smallskip

The vector field generating our large diffeomorphisms is 
\begin{equation}
\zeta_T = \frac{1}{\i \,n}  \left( \partial_t \Phi_T \,\pdv{y}- \partial_y\Phi_T\, \pdv{t}  + \i\, ((y-1)\, \partial_y -1)\Phi_T\, \pdv{\varphi} \right) ,
\end{equation}
with 
\begin{equation}
\Phi_T(y,\tau) = f_1(y)\, e^{\i n \tau}\,.
\end{equation}
Note that the vector field is non-normalizable. It acts non-trivially on the boundary, and thus should not to be treated as pure gauge modes. To see this, we note that near the conformal boundary of \AdS{2} these fluctuations are generated by diffeomorphisms along the vector field 
\begin{equation}
\zeta_T \sim - \xi(\tau) \, \partial_\tau + \xi'(\tau) \,y \,  \partial_y\,, 
\qquad 
\xi(\tau) = \sum_{n}\, \xi_n \,e^{\i n \tau}.
\end{equation}
Therefore, these modes act as time reparametrizations $\tau \to \tau - \varepsilon(\tau)$ while the $y$ or radial component ensures that the boundary curve is of fixed length~\cite{Maldacena:2016upp}. The integration space is 
\begin{equation}
\mathcal{M}_\text{Sch}={\rm Diff}(\mathbf{S}^1)/{\rm PSL}(2,\mathbb{R})\,.
\end{equation}
The quotient is by the isometries of \AdS{2}; curves related by such transformations should not be over counted. In the space of small fluctuations displayed in~\eqref{eq:zeromodes}, the generators of the isometries correspond to $n=-1,0,1$. Consequently,  for we only integrate over $|n|\geq 2$ modes. 

As determined above, at exact extremality with $\Ts=0$, these modes have vanishing action by virtue of being exact zero modes. In JT gravity this corresponds to working on a background with a constant (spacetime independent) dilaton.  

If we focus on the asymptotia, these large diffeomorphisms are solely determined by their behavior near the conformal boundary of \AdS{2}. In writing~\eqref{eq:zeromodes} or~\eqref{eq:zetaT} we have picked a particular way to propagate the modes to the interior of the NHEK geometry. This is determined implicitly by the way that diffeomorphism invariance was fixed in the calculation. One way to implement this is to  work with a gauge fixing term for fluctuations, demanding that they satisfy the harmonic gauge condition.  Demanding that the large diffeomorphisms are zero modes in the presence of this gauge-fixing term imposes the precise form of~\eqref{eq:zeromodes}. One can indeed alternately have determined the modes by solving for the harmonic gauge fixing term $I_\text{gf}$, since any pure diffeomorphism (with a normalizable metric perturbation) would have left $I_{_\text{EH}}$ invariant.

%~~~~~~~~~~~~~~~~~~~~~~~~~~~~~~~~~~~~~~~~~~~~~~~
\subsubsection{The rotational zero mode: a puzzle}\label{sec:rotatez}
%~~~~~~~~~~~~~~~~~~~~~~~~~~~~~~~~~~~~~~~~~~~~~~

On general grounds, it is expected that in addition to these Schwarzian space of zero modes, there is a  second set of zero modes associated to fluctuations in the angular velocity of the black hole. One way to motivate them, as for instance argued in~\cite{Sen:2012cj}, is to consider dimensionally reducing to two dimensions. Since the 4d solution is only axisymmetric, the reduction gives a Maxwell gauge field~\eqref{eq:2dA}. In the language of~\cref{sec:2dJT}, where we describe the NHEK geometry as a two-dimensional dilaton gravity coupled to a ${\rm U}(1)$ gauge field and to matter, these are large gauge transformations acting on the ${\rm U}(1)$ sector. Consequently, we anticipate that the vector fields in \AdS{2} admit a set of zero modes associated with a large gauge transformation of the form $A \to A + \d H$ with 
\begin{equation}
H(y,\tau) =\left( \frac{y-1}{y+1}\right)^{\frac{n}{2}} \left( \delta\lambda_+ \, e^{\i n \tau} + \delta\lambda_-\,  e^{- \i n \tau}\right).
\end{equation}
Again, these zero modes are determined by their behavior at large $y$, and  their profile in the interior should be fixed by our gauge fixing term. In the two-dimensional dilaton gravity theory, it is natural to use a Lorenz gauge fixing term for the ${\rm U}(1)$ gauge field. One can then check that this then is a zero mode with these choices. Once this is determined, we can parametrize these modes by $\delta\lambda_+ \propto v_n $ and the reality condition implies that $\delta\lambda_- \propto \overline{v}_n$. The moduli space of these zero modes is  denoted as $\mathcal{M}_{{\rm U}(1)}$,  but its  details depend on the choice of the ensemble. For example, in a fixed angular velocity ensemble, it is given by $\mathcal{M}_{{\rm U}(1)} = {\rm Loop}({\rm U}(1))/{\rm U}(1)$. 

However, part of our motivation was to not rely on the dimensional reduction, but to rather work directly in the NHEK spacetime in four dimensions. To this end, we should examine if we can find a metric perturbation that could correspond to such a gauge transformation upon reduction. We realize that one can achieve the desired effect by a diffeomorphism $\d\varphi \to \d \varphi + \d H$ or equivalently by a vector field
\begin{equation}\label{eq:xivector}
\xi_R^\mu = H(y,\tau)\,( \partial_\varphi )^\mu\,. 
\end{equation}	

Carrying out this exercise, we find that the metric perturbation generated by $\xi_R^\mu$, viz.,  $h_{\mu\nu} =2 \nabla_{(\mu} \xi^R_{\nu)}$, fails to be annihilated by the Lichnerowicz operator $\Delta_L$. In fact, 
\begin{equation}
\Delta_L h_{\mu\nu} \neq 0 \,, \qquad 
\sqrt{g}\frac{1}{4}\,  \widetilde{h}^{\mu\nu} \Delta_L\, h_{\mu\nu} = \frac{64\,\,n^2}{y^2-1} \,\frac{\sin^5\theta}{(1+\cos^2\theta)^4} \left( \frac{y-1}{y+1} \right)^n \, \delta \lambda_+\, 
\delta \lambda_-\,.
\end{equation}
Tracing through the calculation, we ascertain that the reason for this failure is that the diffeomorphism fails   
to satisfy the harmonic gauge condition. This is indeed the only way for it to have failed, for $I_{_\text{EH}}$ would be unchanged by any diffeomorphism resulting in a normalizable metric perturbation.

One attempt to fix this would be by modifying $\xi_R^\mu$ to bring the perturbation back into the harmonic gauge. We can look for $\xi_H^\mu$ such that $\xi^\mu_{R,H} = \xi_R^\mu  + \xi_H^\mu$ has the same asymptotic behaviour, but the resulting metric perturbation does satisfy the harmonic gauge condition. This again turns out to be tractable, but the resulting vector field is no longer smooth. An example of an explicit solution is
\begin{equation}
\xi^H_\mu  = \left[  c_1 + c_2 \left( \cos\theta+2\,\log\tan\frac{\theta}{2} \right) -4\, \log \sin\theta\right] \bigg(\partial_y H(y,\tau)\, \d y + \partial_\tau H(y,\tau)\, \d \tau \bigg) .  
\end{equation}	
Here $c_1$ and $c_2$ are integration constants, which can be picked at will. Clearly, however, there is no choice of them that ensures that the vector field $\xi^\mu_{R,H}$ is regular at the poles ($\theta = 0,\pi$) of the deformed $\mathbf{S}^2$. This solution is somewhat reminiscent of the singular modes recently obtained in~\cite{Castro:2021csm}.   We explain how we arrive at this result in~\cref{sec:rotApp}, where we rule out a number of other possible ansatz for the compensating diffeomorphism.

\bigskip

At this point, we find ourselves at an impasse. On the one hand, there does not appear to be a metric fluctuation that is in the kernel of the Lichnerowicz operator, and has the right structure to be called a rotational zero mode. In fact, our attempts to solve for a general diffeomorphism associated to a zero mode has not yielded any physically interesting solutions. Unfortunately, we are also unable to provide a proof that no such mode exists. As a piece of circumstantial evidence we note that a similar exercise in the near-extremal BTZ geometry also fails to unearth a rotational zero mode. In that case, we can argue that the absence of the zero mode is consistent with the expectations of the dual CFT. 

On the other hand, we have identified a putative zero mode, albeit one generated by a diffeomorphism that is singular at the poles of the sphere. Even if this were not physically acceptable, one could use the intuition from 2d dilaton gravity coupled to matter, to argue for a different gauge fixing term for the rotational zero mode, viz., one obtained by uplifting Lorenz gauge condition of the ${\rm U}(1)$ gauge field to one for the 4d metric. This choice would break spacetime covariance, as we would impose different gauge choices for the Schwarzian and rotational modes. 

Clearly, neither of these is satisfactory. For now, we will proceed with the rest of the analysis, deferring a more detailed discussion to~\cref{sec:concl}.

%~~~~~~~~~~~~~~~~~~~~~~~~~~~~~~~~~~~~~~~~~~~~~~~
\subsubsection{Incorporating the zero modes}\label{sec:zminc}
%~~~~~~~~~~~~~~~~~~~~~~~~~~~~~~~~~~~~~~~~~~~~~~

We are now ready to write a formal expression for the complete one-loop determinant around the extremal Kerr geometry, after incorporating these two sets of zero modes. The contribution is of two types. The first, comes from the fact that a diffeomorphism produces a change in the metric of order $h_{\mu\nu} \sim \mathcal{O}({\sf J}^2)$. Therefore, we need to compute the Jacobian between the integration measure over metric fluctuations, and the measure over the large diffeomorphisms themselves. This Jacobian factor was carefully  computed by Sen in~\cite{Sen:2012cj}. It was found that each gravitational zero mode\footnote{Take for example the Schwarzian modes. There are an infinite number of them. Nevertheless, the important quantity in the one-loop determinant is the zeta-function regularized number of zero modes, which is $-3$. A similar calculation for the rotational zero modes gives $-1$.}  contributes a factor of $\frac{1}{2} \log S_0$ to  $\log Z$. Including for a moment, both the Schwarzian and the rotational zero modes, we would obtain a total contribution of $-2\log S_0$. In addition, we have a remaining integral over such large diffeomorphisms that still needs to be performed. 

The total one-loop determinant, accounting for the Schwarzian and rotational zero modes, would therefore update~\eqref{eq:logZp} to
\begin{equation}\label{eq:sksk}
\log Z_{\rm 1-loop} =\frac{1}{180} \left( 2 \,n_S - 26 \,n_V + 7 \,n_F + 64\right) \log S_0 + \log \underbrace{ \int_{\mathcal{M}_{\rm Sch}} \d \xi  \,\,\int_{\mathcal{M}_{{\rm U}(1)}} \d v}_{\rm Large~diffeos} \,.
 \end{equation}
Ignoring the matter fluctuations and the integral over the large diffeomorphisms, we reproduce the result quoted 
in~\eqref{eq:senlog}. The previous work on this problem such as~\cite{Sen:2012cj} stops here. The  assumption was that the final term on the r.h.s.\ is independent of $\Js$ and $\Ts$. This is not a well-defined procedure, since the moduli space of zero modes is non-compact and would formally lead to divergent answers. Therefore, these zero modes need regulating. The physical way to do this, proposed in~\cite{Iliesiu:2022onk}, is to turn on a small temperature  and use it as an IR regulator. Specifically, one works with a background that is not just NHEK, but includes small $\order{\Ts}$ corrections. Then we can compute the action over the zero modes and take $\Ts \to 0$ limit at the end of the calculation. We will tackle this in the next subsection.

However, given our issues with the rotational zero mode, if we were to only include the Schwarzian zero modes, then  we would rewrite~\eqref{eq:sksk} as
\begin{equation}\label{eq:skskB}
\log Z_{\rm 1-loop} =\frac{1}{180} \left( 2 \,n_S - 26 \,n_V + 7 \,n_F + 154\right) \log S_0 + \log \underbrace{ \int_{\mathcal{M}_{\rm Sch}} \d \xi  }_{\rm Large~diffeos} \,.
 \end{equation}
 As we will discuss below, in the canonical ensemble the functional integral over the rotational zero mode has no effect. The main change is the coefficient of the $\log S_0$ contribution. In any event, we will simply evaluate~\eqref{eq:skskB} in what follows. This will suffice to demonstrate that the remaining functional integral over the large diffeomorphisms gives us the suppression of density of states at low temperatures.
 
%~~~~~~~~~~~~~~~~~~~~~~~~~~~~~~~~~~~~~~~~~~~~~~~~~
\subsection{Regularization of zero modes and near-extremal Kerr}
%~~~~~~~~~~~~~~~~~~~~~~~~~~~~~~~~~~~~~~~~~~~~~~~~~

Following~\cite{Iliesiu:2022onk}, we first obtain the corrected background around which we re-evaluate the one-loop determinants. The procedure consists in keeping $\Ts$ non-zero but small, and including the leading correction to the NHEK limit. Solving the equation for the temperature in the small $\Ts$ limit, we find the parameters $a$ and $r_+$ behave as 
\begin{equation}\label{eq:MalowT}
\begin{split}
a 
&=
	\Js - \Js^3\, \Ts^2 -4\, \Js^4\, \Ts^3 + \cdots \,, \\
r_+
&=
	\Js + 2\,\Js^2\, \Ts + 5\, \Js^3\, \Ts^2+ \cdots \,. \\	
\end{split}
\end{equation}	

To motivate the correction to the metric at leading order, similar to the derivation of the NHEK geometry in~\cref{sec:LowT}, let us define 
\begin{equation}\label{eq:horexp}
r -r_+ = 2\,\Js^2 \,\Ts \, (y-1)\,, \qquad \tau = \frac{\tE}{\Ts}\,.
\end{equation}	
Proceeding with this ansatz, and taking $\Ts \to0$ first, we find the metric functions reduce to 
\begin{equation}\label{eq:horfnsexp}
\begin{split}
\Delta(r) 
&= 
	4\,\Js^4\, \Ts^2 \left[  (y^2-1) + 4\,\Js \, \Ts\, (y-1) + \cdots \right] , \\
\rho^2(r,\theta)
&=
	\Js^2 \left[ 1+ \cos^2\theta + 4\,\Js\,\Ts\, y + \cdots \right] , \\
\Xi(r,\theta)
&=
	4\,\Js^4 \left[ 
	1 + 4\, \Js \, \Ts\, y + \cdots \right].	
\end{split}
\end{equation}	
Plugging this into the line element and retaining terms to linear order in $\Ts$ (note that so far we have made no assumption about the range of $y$), we find the NHEK geometry and its leading temperature correction
\begin{equation}\label{eq:dsq01alt}
	\begin{split}
	\Js^{-2}\, \d s_0^2 
	&= 	
		g_1 \left[ (y^2-1)\, \d \tau^2  + \frac{\d y^2}{y^2-1} + \d \theta^2 \right] + 4\, g_2 \left( \d \varphi + \i\, (y-1)\, \d \tau \right)^2 , \\ 
	\Js^{-2}\, \d s_1^2 
	&= 	
		\Js\, \Ts \bigg( 
			g_1 \left[ (y-1)\,\left(1-(y^2+y-1) \,c\right) \d \tau^2  + \frac{(y^2+y-1-c)\, \d y^2}{(y^2-1)\,(y+1)} + y\, \d \theta^2 \right] \\
	&	+\, g_2 \left( \d \varphi + \i\, (y-1)\, \d \tau \right)
		\left[ \frac{4\,c}{g_1} \, y\, \left( \d \varphi + \i\, (y-1)\, \d \tau \right) -\i\, (y^2-1)\, (1-c)\, \d \tau \right] 	
		\bigg) .
	\end{split}
\end{equation}
To simplify the expressions we defined the functions of the polar angle $\theta$
\begin{equation}\label{eq:anglefns}
c(\theta) = \cos^2\theta\,, \qquad g_1(\theta) = 1+ c(\theta)\,, \qquad g_2(\theta) = \frac{1-c(\theta)}{1+c(\theta)} \,.
\end{equation}	
Note that the leading temperature correction has a different dependence on the polar angle. In particular, the line element $\d s_1^2$ does not conform to the ansatz used for the two-dimensional reduction~\eqref{eq:ansatz4to2}.

Given this, following~\cite{Iliesiu:2022onk}, we should evaluate the one-loop determinant around a background metric that includes the finite temperature correction computed above, viz., compute the determinant around
\begin{equation}
\d s^2_{\rm background} = \d s^2_0 + \d s^2_1.
\end{equation}
Further corrections are subleading in the low temperature expansion. 

Incorporating the correction at linear order in temperature correction to the background is an easy task for the evaluation of the one-loop determinant of non-zero modes --  the result agrees with~\eqref{eq:logZp} up to corrections that are subleading in the small $\Ts$ limit. Therefore, we do not need to worry about repeating that calculation and  can simply borrow the result. 
The situation is completely different for the modes that have vanishing action at $\Ts=0$, they now have an action of order $\Ts$, which we determine next.

We begin with the Schwarzian modes. We normalize the Schwarzian modes given in~\eqref{eq:zeromodes} as determined in~\cite{Camporesi:1994ga}
\begin{equation}\label{eq:delpmvals}
\delta_+ = \sqrt{\frac{n(n^2-1)}{2\pi}}\, \xi_n\,, 
\qquad
\delta_- = \sqrt{\frac{n(n^2-1)}{2\pi}}\, \xi_n^* \,.
\end{equation}	
This is also the normalization used originally in~\cite{Sen:2012cj}, but differs from that  used in~\cite{Moitra:2021uiv}. The latter work fixes the coefficients $\delta_\pm$ to unity (up to the amplitude factor), and instead incorporates suitable factors in the integration measure that they compute. These choices do not modify the final answer at the level of approximation we are working. They only affect the order one contribution to the entropy, viz., contributions that do not grow with the horizon area or  the temperature. 

We can now compute the on-shell action for the Schwarzian modes, by evaluating~\eqref{eq:EHaction} at quadratic order in $\delta_\pm$ from the total metric $\d s^2_0+\d s^2_1 + \d s^2_\epsilon$. The calculation, while straightforward, still  requires a subtle treatment of counterterms and boundary terms related to the choice of ensemble. Relegating the details of this procedure to~\cref{app:schwaction}, we find at the end of the day, for a single mode of a given value of $n$, an action to quadratic order for the Schwarzian modes:
\begin{equation}\label{eq:SosT}
I_{_\text{EH}} = \frac{2\pi\, \Js^2}{G_N} + \frac{4\pi\, \Js^3\, \Ts}{G_N} - \frac{4\pi\, \Js^3\, \Ts}{G_N}\, \frac{\delta_+\, \delta_-}{4\,(n^2-1)}\,.
\end{equation}	
Using the given values of $\delta_\pm$ in~\eqref{eq:delpmvals} and including the sum over all Fourier modes, we end up with 
\begin{equation}
I_{_\text{EH}} = \frac{2\pi\, \Js^2}{G_N} + \frac{4\pi\, \Js^3\, \Ts}{G_N} -  \frac{4 \pi^2\Ts}{\Ts_q}\,\sum_{|n|>1} \frac{n}{8\pi}\, \xi_n\, \xi_n^*\,.
\end{equation}	

As advertised, the Schwarzian modes have been lifted with the leading temperature correction. This is the same as the result obtained in the spherically symmetric solutions examined hitherto~\cite{Iliesiu:2020qvm,Moitra:2021uiv,Iliesiu:2022onk}, despite the non-trivial dependence on the polar angle in the temperature corrected background.

Evaluating the one-loop determinant around the near-extremal Kerr geometry gives a small action to these modes, of order ${\sf J}^3 \Ts$. Now the evaluation of the path integral over these Schwarzian modes is well-defined. Using zeta-function regularization these modes lead to a contribution
\begin{equation}\label{eq:Zsch1lp}
\log Z_\text{Sch} = \log \prod_{n=2}^\infty \frac{ 2G_N}{\Js^3\,\Ts\, n} = \frac{3}{2} \log \frac{\Ts}{\Ts_q} + \order{1} \,.
\end{equation}	
We emphasize that the order one term in the r.h.s.\ is not reliable. For example, even the non-zero modes would also contribute at this order, and we have not included them in our temperature corrected analysis. 

Finally,  let us comment on the rotational modes. Even if they were genuine zero modes of the NHEK background, we expect them to have an action of order $\Ts$ in the fixed angular velocity ensemble. Their quantization nevertheless is simpler. Since ${\rm U}(1)$ is compact, the quantum mechanics of these modes are gapped. In particular, if we work in a fixed charge sector, viz., fixing the angular momentum in our case, then their contribution can only at most shift the ground state energy. 

%~~~~~~~~~~~~~~~~~~~~~~~~~~~~~~~~~~~~~~~~
\subsection{The entropy of near-extremal Kerr}
%~~~~~~~~~~~~~~~~~~~~~~~~~~~~~~~~~~~~~~~~

We are now ready to put all the pieces together and compute the final form of the free energy of the Kerr black hole, in an ensemble of fixed angular momentum. The first ingredient is the on-shell classical action on the near-extremal (finite but small $\Ts$) background. It is given by 

\begin{equation}
\log Z_{\rm tree} = S_0 + 2\pi^2 \frac{T}{T_q} +\mathcal{O}(T^2)\,.
\end{equation}
We keep the leading order correction at finite temperature. The second ingredient is the one-loop determinant, including the Schwarzian mode contribution. The answer, combining~\eqref{eq:Zsch1lp} and~\eqref{eq:skskB}, is given by
\begin{equation}
\log Z_{\rm 1-loop} =\frac{1}{180} \left( 2 \,n_S - 26 \,n_V + 7 \,n_F + 154\right) \log S_0 + \frac{3}{2}\, \log \frac{T}{T_q}+\order{1}.
\end{equation}
We emphasize again that the order one correction is not reliable since we have not kept track of it in the heat kernel expansion. Putting these two results together, we obtain the total free energy 
\begin{equation}
-\beta F = S_0 + 2\pi^2 \frac{T}{T_q} + \frac{1}{180} \left( 2 n_S - 26 n_V + 7 n_F + 154\right) \log S_0 + \frac{3}{2}\, \log \frac{T}{T_q}+\order{1}\,.
\end{equation}

From this expression we can compute the entropy as a function of temperature using thermodynamic identities. We obtain the final answer,
\begin{equation}
S= S_0  + \frac{1}{180} \left( 2 \,n_S - 26 \,n_V + 7 \,n_F + 154\right) \log S_0 + 4\pi^2 \frac{T}{T_q}+ \frac{3}{2}\, \log \frac{T}{T_q}+\order{1}.
\end{equation}
This is our prediction for the quantum entropy of black hole microstates of the Kerr black hole. For the Standard Model the light fields relevant to the one-loop determinant at energies lower than electroweak scale, and for black holes smaller than the Hubble scale, is the photon and three neutrinos, and $n_V=1$, $n_F=3/2$ leading to 
\begin{equation}
    S_{\rm SM} =  S_0  + \frac{277}{360} \log S_0 + 4\pi^2 \frac{T}{T_q}+ \frac{3}{2}\, \log \frac{T}{T_q}+\order{1}\,.
\end{equation}
On the other hand, for pure gravity $n_S=n_V=n_F=0$, and thus we arrive at the result quoted in~\cref{sec:intro}, viz.,
\begin{equation}
S_{\rm pure~grav} = S_0  + \frac{154}{180} \log S_0 + 4\pi^2 \frac{T}{T_q}+ \frac{3}{2}\, \log \frac{T}{T_q}+\order{1}\,.
\end{equation}
This calculation does not take into account instabilities of the Kerr black hole in flat space that would appear as imaginary components to $\log Z$ that determine the lifetime of the black hole. We comment on this in~\cref{sec:concl}.

%~~~~~~~~~~~~~~~~~~~~~~~~~~~~~~~~~~~~~~~~~~~~~~~~~
\section{Generalizations}\label{sec:genralizations}
%~~~~~~~~~~~~~~~~~~~~~~~~~~~~~~~~~~~~~~~~~~~~~~~~~

In this section we will generalize our analysis to more complicated near-extremal rotating black hole. We first consider in~\cref{sec:KerrNewman} the extremal limit of a charged rotating black hole in asymptotically flat space. Next in~\cref{sec:KerrAdS} we analyze an uncharged rotating black hole in asymptotically \AdS{4}. Even though the results are quantitatively different, we find qualitatively the same behavior as the extremal limit of the Kerr black hole.

%~~~~~~~~~~~~~~~~~~~~~~~~~~~~~~~~~~~~~~~~~~~~
\subsection{The extremal Kerr-Newman black hole}\label{sec:KerrNewman}
%~~~~~~~~~~~~~~~~~~~~~~~~~~~~~~~~~~~~~~~~~~~~

In this section we consider near-extremal black hole geometries which are solutions to a gravity theory including at least an Einstein-Maxwell sector
\begin{equation}\label{eq:EMaction}
I_{_\text{EM}}
= 
	- \frac{1}{16 \pi G_N}  \int_\mathcal{M} \dd[4] x\, \sqrt{g} \left[ R - F_{\mu\nu}F^{\mu\nu} \right] + I_{\rm GHY},
\end{equation}
where $F_{\mu\nu} = \partial_\mu A_\nu - \partial_\nu A_\mu$ and $A_\mu $ is a ${\rm U}(1)$ Maxwell potential. When imposing Dirichlet boundary conditions on all fields, the boundary terms include the Gibbons-Hawking-York term and also an electromagnetic boundary term $\oint \d\Sigma^\mu F_{\mu\nu} A^\nu$~\cite{Braden:1990hw, Hawking:1995ap}. The black hole solution of this theory is the Kerr-Newman black hole. We work instead in an ensemble where we fix $J$, the angular momentum, and $Q$, the electric charge. Following a similar notation as in the uncharged case, we label the black hole by the parameter $a= J/M$ and by the location of the outer horizon which at extremality becomes $r_+ = \sqrt{Q^2 + a^2}$.

To compute the near-extremal entropy of Kerr-Newman black hole we follow a similar procedure as in the previous section. First, we can take the zero temperature limit after performing a change of coordinates as in~\eqref{eq:NHEKm2} to zoom into the near horizon region. The generalization of the NHEK geometry when charge is present is
\begin{equation}\label{eq:NHEKN}
 \dd s_0^2 
= \Gamma(\theta)\left[ (y^2-1) \dd \tau^2 + \frac{ \dd y^2}{y^2-1} +  \dd \theta^2+\gamma(\theta)^2 ( \dd \varphi +\ii \,k \,(y-1) \,\dd \tau)^2\right] .
\end{equation}
where we define the functions
\begin{equation}
\Gamma(\theta) = r_+^2 + a^2 \cos^2 \theta\,, \qquad 
\gamma(\theta) = \frac{(r_+^2 + a^2)\sin \theta}{r_+^2 + a^2 \cos^2 \theta}\,, \qquad k = \frac{2 a r_+}{r_+^2 + a^2} \,.
\end{equation}
These functions implicitly depend on $(Q,J)$ through the variables $(r_+,a)$. Finally, we also need to perform the same transformation of the gauge potential, which becomes
\begin{equation}
A_\mu\,\d x^\mu = f(\theta) \, (\d \varphi + \i\, k\, (y-1) \d \tau) - \frac{e}{k}\, \d \varphi\,,
\end{equation}
where 
\begin{equation}
    f(\theta) = Q \frac{r_+^2 + a^2}{2 r_+ a} \, \frac{r_+^2 - a^2 \cos^2\theta}{r_+^2 + a^2 \cos^2 \theta}
    \,, \qquad 
    \qquad e = \frac{Q^2 }{r_+^2 + a^2}.
\end{equation}

The extremal horizon area of this black hole, in our conventions, is given by 
\begin{equation}
S_0 = \frac{A_{\rm extremal}}{4 G_N} = \frac{\pi\, (2 a^2 + Q^2)}{G_N} .
\end{equation}
The extremal mass is $M_0=\sqrt{a^2 + Q^2}$ and $a= J/M_0$. The one-loop determinant around this geometry was computed in~\cite{Bhattacharyya:2012wz}. The calculation is similar to~\cite{Sen:2012cj}; it uses the Seeley-de Witt expansion to evaluate the temperature independent correction, and ignores the temperature dependent effects from zero modes. To apply this formalism one needs to take $Q$ to be large and work with a fixed ratio 
\begin{equation}
Q \to \infty\,, \qquad b= a/Q ~{\rm fixed}.
\end{equation}
That way $Q$ represents the overall scale of the metric and sources logarithmic corrections computed by the Seeley-de Witt expansion.

We want to correct this last step described in the previous paragraph, just as done for Kerr above. We will mention some salient features compared to the previous treatment, in particular, mainly focusing on the contribution from the gravitational mode. Once again we consider gauge fixing to the harmonic gauge, including the same $I_\text{gf}$ contribution~\eqref{eq:gaugefix} to the action~\eqref{eq:EMaction}. We have to keep track of the metric dependence from the Maxwell term. Including this, the kernel of the quadratic action for metric fluctuations $h_{\mu\nu}$ becomes
\begin{equation}
    \Delta_{L}^{\rm EM} h_{\mu\nu} = \Delta_L h_{\mu\nu} + \frac{1}{2} F^2 g_{\mu\nu} h^{\rho}{}_\rho - F^2 h_{\mu\nu} + 4 F_{(\mu\rho}F_{\nu \sigma)} h^{\rho \sigma}+ 8 F_{(\mu\rho} F_{\sigma}{}^{\rho} h^\sigma{}_{\nu)}+4 F_{(\mu\rho} F_{\nu)}{}^\rho h^\sigma{}_\sigma,
\end{equation}
where $F^2 = F_{\mu\nu} F^{\mu\nu}$ and the parenthesis between indices indicates symmetrization of the pair $T_{(ab)} = \frac{1}{2} (T_{ab} + T_{ba})$. The superscript EM indicates this kernel applies to a solution of Einstein-Maxwell equations. 

Similar to the uncharged case, we find a robust Schwarzian zero mode labeled by an integer $n\in \mathbb{Z}$ different from $n\neq 0,\pm1$. We can construct it by implementing a large diffeomorphism
\begin{equation}\label{eq:zetaTkn}
\zeta_T = e^{\i n \tau} \left(f_1(y)\, \pdv{y} + f_2(y) \pdv{t} + f_3(y) \pdv{\varphi} \right), 
\end{equation}
where now the functions become
\begin{equation}
\begin{split}
&f_2(y) =  \frac{\i\, f_1'(y)}{n}\,,\qquad  f_3(y) = \frac{2a r_+ f_1'(y) (y-1) - 2 a r_+ f_1 (y)}{n (r_+^2 + a^2)} \,,\\
& f_1(y) =\left(\frac{y-1}{y+1} \right)^\frac{n}{2}\,(n+y) \,\frac{\delta_+}{2\,(n^2-1)} . 
\end{split}
\end{equation}
The change in the metric generated by this diffeomorphism is an obvious generalization of~\eqref{eq:zeromodes}, namely
\begin{equation}
\d s_\text{Sch}^2 
= \Gamma(\theta) \left[ H_{\tau \tau}(\tau,y) \, \d \tau^2 + 2\, H_{\tau y}(\tau, y) \, \d \tau \,\d y + H_{yy}(\tau, y)\,  \d y^2 \right] , 
\end{equation}
and the functions $H_{\mu\nu}(\tau,y)$ are the same that appeared in~\eqref{eq:zeromodes}. One can check by explicit calculation that 
\begin{equation}
\Delta_L^{\rm EM} \nabla_{(\mu} \zeta_T{}_{\nu)} = 0, 
\end{equation}
making it an exact zero mode of the action. 

To treat this zero mode we will raise the temperature of the black hole, which amounts to keeping the first correction to the low temperature limit that produced~\eqref{eq:NHEKN}. This gives an action to this mode proportional to $T/T_q$ and produces a $\frac{3}{2} \log \frac{T}{T_q}$ correction to the entropy. $T_q$ is now a generalization of the quantity introduced in~\cref{sec:LowT}, that now also depends on the charge, and can be extracted from the classical term in entropy that is linear in $T$. It is given by
\begin{equation}
T_q = \frac{\pi}{G_N \,M_0\, S_0}\,.
\end{equation}
This relation is valid for any charge and angular momentum, see~\cite{Turiaci:2023wrh}. Therefore, quantum effects from the Schwarzian mode become large for temperatures smaller than this scale.

The other expected zero mode is the rotational mode which we can write again as
\begin{equation}\label{eq:xivectorKN}
\xi_R^\mu = H(y,\tau)\,( \partial_\varphi )^\mu\,. 
\end{equation}
Just as for Kerr, we find this mode has a non-vanishing action since
\begin{equation}
\Delta^{\rm EM}_L \nabla_{(\mu} \xi^R_{\nu)} \neq 0.
\end{equation}
More specifically, the right hand side does vanish when $a=0$, and it is of order $a$ when $a\ll r_+$. Our calculations are consistent with the fact that we do find a rotational zero mode in Reissner-Nordstrom that is actually enhanced to a set of ${\rm SU}(2)$ modes. We have not been able to find a compensating diffeomorphism that puts this mode in our preferred gauge. Therefore, the same puzzle we find for $Q=0$ persists at all charges (unless $a=0$ as mentioned above).

To summarize, we find the near-extremal entropy of the Kerr-Newman black hole to be
\begin{equation}
S_{\rm Kerr-Newman} (T) = S_0 + f(b) \log S_0 + \frac{3}{2} \log \frac{T}{T_q},
\end{equation}
The first term is the classical Bekenstein-Hawking area, the second term comes from the temperature independent contribution to the heat kernel, including the zero mode correction, and the last term comes from a proper treatment of the Schwarzian mode. The temperature independent contribution has two pieces $f(b) = f_0 + f_{hk}(b)$: one from the zero modes, $f_0$, and another,  $f_{hk}$, from the direct heat kernel calculation ignoring zero modes. These are in turn
\begin{equation}
    \begin{split}
        f_0 &= -\frac{3}{2}\,, \\
        f_{hk} &= \frac{1233\, (2\,b^2+1)^2 \tan^{-1}\left(\frac{b}{\sqrt{b^2+1}}\right) - 
        b\,\sqrt{b^2+1}(-463+3080 \,b^2 + 7960\, b^4 + 3184 \,b^6)}{720\,b\,(b^2+1)^{5/2}\, (2\,b^2+1)} .
    \end{split}
\end{equation}
Due to the fact that we did not find a rotational zero mode for $b\neq 0$ our result for $f(b)$ differs from the result in~\cite{Bhattacharyya:2012wz}. According to~\cite{Bhattacharyya:2012wz} the zero mode contribution is $f_0=-2$ since they argue for a total of $-4$ zero modes. This is the same puzzle raised for Kerr in~\cref{sec:rotatez}. For $b=0$ we recover the Reissner-Nordstrom result found in~\cite{Sen:2011ba} and~\cite{Iliesiu:2022onk}, and the puzzle disappears.

%~~~~~~~~~~~~~~~~~~~~~~~~~~~~~~~~~~~~~~~~~~~~
\subsection{The Kerr-AdS black hole in four dimensions}\label{sec:KerrAdS}
%~~~~~~~~~~~~~~~~~~~~~~~~~~~~~~~~~~~~~~~~~~~~

A second generalization we can consider is the rotating \AdS{4} black hole~\cite{Carter:1968ks}, which for simplicity we dub the Kerr-AdS black hole.\footnote{This is an interesting case to consider since in some situations BPS rotating (charged) black holes in AdS in four and five dimensions can be matched with the boundary CFT results, such as~\cite{Cabo-Bizet:2018ehj,Choi:2018hmj,Benini:2018ywd}, particularly in light of~\cite{Boruch:2022tno}.} In standard Boyer-Lindquist coordinates takes the form:
\begin{equation}\label{eq:kerrads4bl}
\begin{split}
\d s^2 &= 
	-\frac{\Delta_r}{\rho^2} \, \left( \d t- \frac{a}{\Xi}\, \sin^2\theta\, \d \varphi\right)^2 +
	 \frac{\rho^2}{\Delta_r}\, \d r^2 + \frac{\rho^2}{\Delta_\theta}\, \d \theta^2 
	 +\frac{\Delta_\theta\, \sin^2\theta}{\rho^2} \, \left(a\, \d t-  \frac{r^2 + a^2}{\Xi}\, \d \varphi\right)^2\,,  \\
\Delta_r(r) &=
	 (r^2+ a^2) \left(1+ \frac{r^2}{\lads^2}\right) -  (r_+^2+ a^2) \left(1+ \frac{r_+^2}{\lads^2}\right)  \frac{r}{r_+}\,,
 \\
\Delta_\theta(\theta) &=
	 1-\frac{a^2}{\lads^2}\, \cos^2\theta \ , \qquad \rho^2(r,\theta) = r^2 +a^2\, \cos^2\theta \ , \qquad \Xi = 1-\frac{a^2}{\lads^2}\,.	 
 \end{split}
\end{equation}
The Euclidean solution is obtained by $t\to \i\,\tE$. The solution is parameterized by two parameters which we have taken to the horizon size $r_+$ and the rotation parameter $a$. The intensive thermodynamic parameters are 
\begin{equation}\label{eq:TOSkerr4}
T 
= 
	\frac{r_+ \,(1+a^2 \, \lads^{-2} + 3\, r_+^2\, \lads^{-2} - a^2\, r_+^{-2})}{4\pi\, (r_+^2 +a^2)}\,, \qquad 
\Omega  
= 
	a\, \frac{ 1+\frac{r_+^2}{\lads^2}}{r_+^2+a^2} \,,
\end{equation}
while the extensive quantities are 
\begin{equation}\label{eq:kerradsMJ}
M = \frac{m}{2\,G_N\, \Xi^2} \,, \qquad J = \frac{m \,a }{2 \,G_N\,\Xi^2} \,, \qquad m  \equiv  \frac{r_+^2+ a^2}{r_+}\left(1+ \frac{r_+^2}{\lads^2}\right) .
\end{equation}	

The extremal Kerr-\AdS{4} solution is attained along a codimension-1 locus in parameter space (setting $T=0$). It is clearly characterized by a single scale, $r_0$, which sets the size of the near-horizon \AdS{2} length scale, viz., 
\begin{equation}
r_+ 
= 
	r_0 \,, \qquad 
a_0 
=
	r_0 \sqrt{\frac{\lads^2+ 3\, r_0^2}{\lads^2-r_0^2}} \,.
\end{equation}
 The line element of this near-horizon solution is of the general form given in ~\eqref{eq:NHEKN}, albeit with different angular dependence, cf.,~\cite{Hartman:2008pb} for the explicit geometry. 

In  the non-extremal solution~\eqref{eq:kerrads4bl} the physical charges~\eqref{eq:kerradsMJ} are extensive, but there is a second length scale set by the background cosmological constant $\lads$, which may not scale homogeneously with the charges. Consequently, in analyzing the logarithmic contribution from the heat kernel~\eqref{eq:logZp}, one will end up with $\log S_0 + \log F(M\lads,J) $, where the function $F$ cannot be determined without detailed knowledge of the eigenspectrum. However, in the near-horizon geometry is characterized by a single macroscopic length scale.  Inspired by this,~\cite{David:2021eoq} analyzes the Seeley-de Witt expansion and extracts the logarithmic correction in the area for rotating \AdS{4} black holes.

Focusing instead on the temperature corrections, we need to only examine the gravitational zero modes. The analysis in the near-horizon geometry proceeds as in the previous examples. Including the contribution from the Schwarzian mode, we predict a low temperature entropy of the form quoted in~\eqref{eq:sensch} with 
\begin{equation}
S_0 = \frac{\pi}{G_N} \, \frac{r_0^2+a_0^2}{1- \tfrac{a_0^2}{\lads^2}} \,, \qquad T_q = \left(1- \frac{a_0^2}{\lads^2}\right)  \frac{\lads^2 + 6\, r_0^2 + a_0^2}{r_0\, \lads^2 \, (r_0^2 + a_0^2)}\,,
\end{equation}
and the $\log S_0$ coefficient as determined in the aforementioned reference.

%~~~~~~~~~~~~~~~~~~~~~~~~~~~~~~~~~~~~~~~~
\section{Conclusions}\label{sec:concl}
%~~~~~~~~~~~~~~~~~~~~~~~~~~~~~~~~~~~~~~~

The main result of the paper is an explicit demonstration that  near-extremal Kerr black holes do not have a macroscopic ground state degeneracy, in contrast to the leading classical expectation. As presaged at the outset, this is indeed what one might expect based on the general arguments given in~\cite{Iliesiu:2020qvm}. The suppression of the classical entropy arises due to geometric zero modes supported in the near-horizon. These modes are to be treated quantum mechanically, and their functional integral induces a power-law suppression at low temperatures. The analysis we undertake is valid for low, but not arbitrarily low, temperatures. In particular, we expect the result~\eqref{eq:sensch} to hold for $T\leq T_q$, but not for temperatures that are exponentially (in $S_0$) below this scale. To ascertain effects in this regime would require control over other saddle point configurations to the quantum gravity path integral, which at the moment one does not have. 

The primary motivation for undertaking the exercise, whose result one might have viewed as a foregone conclusion, was to test the general principles directly in four dimensional gravity, without reducing to two-dimensional dilaton gravity. At a technical level we are led to examining the kernel of the spin-2 fluctuation operator about the near-horizon of the extremal Kerr geometry. This turns out to be a feasible, though non-trivial, exercise thanks in part to the enhanced symmetry. Curiously, we only find three zero modes in the NHEK geometry. These Schwarzian modes would correspond in the 2d dilaton gravity description of~\cref{sec:2dJT} to \AdS{2} boundary graviton modes. 

Naively, one also expects another zero mode associated with the fluctuation of the angular velocity. From the perspective of~\cref{sec:2dJT} this is associated with the large gauge transformation of the Maxwell field, arising from the fibration structure. Note that in a spherically symmetric solution, one would have obtained a $\mathrm{SO}(3)$ gauge field, but the axisymmetric background only has a $\mathrm{U}(1)$. The putative zero mode which can be identified using the result of~\cite{Camporesi:1994ga} satisfies the Lorenz  gauge condition in 2d. However, when lifted to the full 4d NHEK geometry we find that it is not annihilated by the spin-2 fluctuation operator. While the mode in question is generated by a diffeomorphism, it violates the gauge condition chosen to simplify the quadratic fluctuations (harmonic gauge). Furthermore, attempting to bring it into our preferred gauge leads to a vector field that is singular at the poles of the $\mathbf{S}^2$. We emphasize that this issue is particular to the rotating solutions; for spherically symmetric black holes we recover the expected zero modes associated with the isometries of the sphere.

We adhere to the philosophy that the physical zero modes should correspond to large diffeomorphisms along smooth vector fields, and that there is nothing strange with the harmonic gauge choice. There are several alternatives one could have considered: allow for the vector field to be singular while staying within the gauge, or make different gauge choices (breaking covariance in the action) for different modes. Either of the alternatives would have given us a rotational zero mode, whose presence a-priori seems sensible on physical grounds. In our final result for the entropy, the presence or absence of this rotational zero mode does not affect the temperature dependent terms, but we are led to a different coefficient for the $\log S_0$ contribution than that ascertained in the earlier analysis of~\cite{Sen:2012cj}. 

To test our techniques, we have attempted to examine the rotating BTZ geometry. This geometry is simpler, the \AdS{2} part is fibered over an $\mathbf{S}^1$ and there is no warping. We have verified the presence of the Schwarzian zero modes, but found no sign of a rotational zero mode. In fact, the absence of the rotational zero mode is consistent with holographic duality. The rotating BTZ solution is dual to a 2d CFT at non-zero temperature. From a CFT analysis one finds that the low temperature correction to the partition function is given by $T^\frac{3}{2}$ in both the canonical and the grand canonical ensemble~\cite{Ghosh:2019rcj}. However, it also bears noting that the chemical potential in this configuration is frozen, so it is natural not to expect a zero mode. We do not have an analogous argument for the Kerr black hole.

The issue with the rotational zero mode is stark if we consider supersymmetric solutions like the five-dimensional BMPV geometry.\footnote{ We thank Ashoke Sen for emphasizing this point to us.} In that case, the semiclassical  path integral computation can be contrasted against a microscopic counting. As demonstrated in~\cite{Sen:2012cj} one has exact agreement between the two approaches vis-\'a-vis the $\log S_0$ correction to the entropy, once one correctly accounts for the zero modes. The count goes as follows: the BMPV solution preserves $\mathrm{SU}(2) \times \mathrm{U}(1)$ isometry. The near-horizon solution (see for instance~\cite{Gupta:2021roy}) is then expected to have 3 Schwarzian zero modes, 3 zero modes from the $\mathrm{SU}(2)$ isometry, and a rotational zero mode for a total of 7 gravitational zero modes.  Indeed, it is only with these 7 zero modes, one does match with the microscopic calculation. The structure of the near-horizon geometry is similar to the BTZ case, and as in that case, we have been unable to identify a suitable candidate for the rotational zero mode (we do recover the remaining six zero modes). For completeness, we note that the solution also has zero modes associated to Maxwell fields sourcing the background (depending on the string theory embedding), which one can find quite directly. To fix this issue one should therefore either explain why the rotational mode is indeed a zero-mode, or find another mode (perhaps fermionic), assumed in \cite{Sen:2012cj} to be a zero mode, to have an action such that the $\log S_0$ is modified in a way that the final answer matches the microsocopic result from string theory. 

The puzzle of the rotational zero mode notwithstanding, one has a clear statement for the low temperature thermodynamics of rotating black holes. For simplicity, we have only described the results for four dimensional black holes in asymptotically flat (and AdS) spacetimes. The analysis can be readily generalized to other dimensions. The only ingredients one needs are the near-horizon region and the leading thermal correction. The former is readily available, thanks to the classification of extremal near-horizon geometries, cf., ~\cite{Kunduri:2013gce} for a review of this program (for known vacuum black holes explicit near-horizon geometries were compiled in~\cite{Figueras:2008qh}).  It would be interesting to complete the zero mode analysis and deduce their lifting at low temperatures. 

We now turn to some other outstanding issues and comment on ideas in the literature for understanding Kerr entropy, the physical interpretation of the zero modes in the near-horizon geometry, and other physical consequences their presence could have. 

\paragraph{Comments on Kerr/CFT:} We recall that in~\cite{Guica:2008mu} it is argued that diffeomorphisms of the NHEK geometry generated by the vector field
\begin{equation}\label{eq:kerrcftV}
\begin{split}
\zeta_\mathfrak{f} 
&= 
	\mathfrak{f} (\varphi)\, \pdv{\varphi} - \mathfrak{f}'(\varphi) \,  \pdv{y} \,,
\end{split}
\end{equation}
leads to an enhancement of the $\mathrm{SL}(2,\mathbb{R}) \times {\rm U}(1)$ isometry to a chiral  Virasoro algebra. The central charge of the Virasoro algebra  is fixed by the angular momentum, $c_L = 12\, J$. Exploiting the fact that the natural ground state has temperature $T_L = \frac{1}{2\pi}$ and using Cardy asymptotics, one reproduces the extremal entropy $S = 2 \pi\, J$. An extension to the near-extremal situation identifies another chiral Virasoro symmetry~\cite{Castro:2009jf} with $c_R = 12\, J$, and recovers the classical correction to the near-extremal entropy (the linear $T$ correction).

There are two observations to be made regarding this proposal in light of our analysis. Firstly, from the perspective of the gravitational path integral the diffeomorphisms~\eqref{eq:kerrcftV} lead to non-normalizable deformations of the \AdS{2} asymptotics. Indeed, implementing a diffeomorphism with some profile function $\mathfrak{f}(\varphi)$ we find
 \begin{equation}
     \delta_{\mathfrak{f}}\,  \d s^2 
= 
	2\, g_1 \left(
	\left[\frac{\d y^2}{(y^2- 1)^2} -y^2\, \d t^2 + \frac{4\,g_2}{g_1}  \left(y (y-1)\, \d t^2 + \d\varphi^2 - \i\, \d t\, \d\varphi \right)\right]  \mathfrak{f}' - \frac{\d y\,\d\varphi}{y^2- 1}\, \mathfrak{f}'' \right) .
 \end{equation}
This asymptotic behaviour was indeed an input in ascertaining which class of diffeomorphisms to admit in~\cite{Guica:2008mu}. However, the vector field~\eqref{eq:kerrcftV} is not distinguished in any way in the analysis of the near-horizon fluctuation spectrum, and the resulting deformation has a divergent on-shell action.

A second perspective is offered by the insight of~\cite{Ghosh:2019rcj} who demonstrate a universal Schwarzian sector in 2d CFTs. Assuming that the CFT has a non-vanishing twist gap (viz., there are no conserved currents), and a large central charge $c\gg 1$, in the canonical ensemble with $T\sim \order{c^{-1}}$ and $J \sim \order{c^3}$, one finds
\begin{equation}\label{eq:cftSchwarz}
    Z_\text{CFT}(T,J) \sim (c\,T)^\frac{3}{2}\, J^{-\frac{3}{4}}\, \exp\left[2\pi\,\sqrt{\frac{c\,J}{6}} + \frac{\pi\,c}{12}\, T - \frac{1}{T}\, \left(J - \frac{1}{12}\right)  \right] .
\end{equation}
This result which follows from the vacuum dominance in a dual channel, can be demonstrated to hold for $T \ll c^{-1}$ rigorously~\cite{Pal:2023cgk}.  This universal high-spin regime is not attained if the central charge scales like the angular momentum. One could attempt to match the result obtained from the path integral computation extrapolating~\eqref{eq:cftSchwarz} past its regime of validity. Doing so would appear to give us the desired result (assuming as earlier that the left movers contribute to the extremal entropy and the right movers to the thermal excitations), but requires further justification.

\paragraph{An asymptotically flat perspective:} In this paper we have computed the quantum corrections to the entropy of near-extremal Kerr black holes. The calculation consisted in an evaluation of the one-loop determinants around the extremal geometry. We found the presence of exact zero modes localized in the NHEK throat. These zero modes need to be quantized. Making this problem well-defined requires turning on a small temperature, and thus going away form the extremal limit. Other than this IR regulator, most the calculation was mainly performed in the NHEK throat. Therefore, it is worth thinking about the role of these zero modes in the full, asymptotically flat, geometry. What are these zero modes in the asymptotically flat extremal solution?

To answer this question one needs to look carefully at the topology of the extremal black hole. While a non-extremal black hole has the topology of a cigar, the extremal black hole bears closer resemblance to a cylinder (although the size of the thermal circle becomes infinite) since the horizon has been pushed infinitely far down the throat. When studying gravitational fluctuations on a space with boundary we need to impose boundary conditions to separate diffeomorphisms that are gauged from those that physically act on the boundary. The choice at the asymptotically flat (or AdS) boundary can be made once and for all at the spacetime asymptopia. Nevertheless, in the extremal background we need to decide how to treat the diffeomorphism which becomes non-vanishing as we approach the horizon. The physical answer is that they should not be gauged. Instead, they can be glued smoothly to the large diffeomorphisms we study in the NHEK throat. Consequently, they should be treated as exact zero modes at zero temperatures, which get a non-vanishing action once temperature is turned on.

\paragraph{Instabilities of the Kerr black hole:} The extremal limit of Kerr or Kerr-Newman black hole suffers from instabilities, which we did not incorporate in this analysis. One example is superradiance\footnote{\textit{Note added:} As mentioned at the end of the introduction the time-scale associated to superradiance is the same as the time scale set by $T_q$. This fact affects the possibility of constructing, even in principle, Kerr black holes close enough to extremality such that the Schwarzian quantum effects could be tested.} in the uncharged case, and another is Schwinger pair production (in addition) in the charged case. A different type of instabilities was recently discovered in~\cite{Horowitz:2023xyl} when higher derivative corrections to the Einstein-Hilbert action are included (such as in string theory). In the presence of matter, the Aretakis instability~\cite{Aretakis:2011ha}, which owes to additional conserved charges at the horizon, is yet another possibility.  Similar issues for persist for the solutions with negative cosmological constant, where there are interesting implications for the dual holographic CFT~\cite{Cardoso:2004hs,Kim:2023sig}.

All these mechanisms take place close to the horizon. They can therefore can be analyzed directly in the NHEK throat  (or generalizations thereof). In particular, it is useful to follow the approach of~\cref{sec:2dJT} and describe the dynamics near the horizon in terms of an \AdS{2} gravity theory. From this vantage point the instabilities can be viewed as arising due the presence of tachyonic matter modes in \AdS{2} coupled to Jackiw-Teitelboim gravity. In this way of thinking about the problem, it does not matter whether the 2d fields arise from a higher dimensional geometric or matter degree of freedom. Some physical applications such as the one proposed in~\cite{Horowitz:2023xyl} (and essentially any of the other instabilities mentioned in the previous paragraph) require being arbitrarily close to extremality and therefore quantum effects from the Schwarzian mode should be taken into account if one wants to work at temperatures smaller than $T_q$. This is an interesting open problem.

%~~~~~~~~~~~~~~~~~~~~~~~~~~~~~~~~
\section*{Acknowledgements}
%~~~~~~~~~~~~~~~~~~~~~~~~~~~~~~~~

It is a pleasure to thank R.~Emparan, C.~Ferko,  M.~Heydeman, V.~Hubeny, L.~Iliesiu, R.~Loganayagam, J.~Maldacena, S.~Murthy, L.~Pando Zayas, A.~Porfyriadis, W.~Song, and S.~Trivedi for useful discussions. We would specifically like to acknowledge important insightful discussions with D.~Marolf and A.~Sen; we are very grateful to them for sharing their thoughts. MR was supported by U.S.\ Department of Energy grant DE-SC0009999 and funds from the University of California. MR is grateful to the long-term workshop YITP-T-23-01 held at YITP, Kyoto University, for hospitality during the concluding stages of this project. GJT was supported by the Institute for Advanced Study and the NSF under Grant No. PHY-2207584, and by the Sivian Fund, and currently by the University of Washington and the DOE award DE-SC0024363.

\appendix
%~~~~~~~~~~~~~~~~~~~~~~~~~~~~~~~~~~~~~~~~
\section{Evaluation of the classical action}\label{app:schwaction}
%~~~~~~~~~~~~~~~~~~~~~~~~~~~~~~~~~~~~~~~~ 

To evaluate the classical contributions to the on-shell action of the Kerr black hole, we can either work in the full solution~\eqref{eq:Kerr} retaining the asymptotically flat region, or zoom into the near-horizon region retaining the leading near-extremal contribution~\eqref{eq:dsq01alt}. We describe briefly both calculations in turn, emphasizing the need to include additional counterterms in the near-horizon region. 

When computing the on-shell action for Kerr solution~\eqref{eq:Kerr} itself directly, there is no bulk contribution, as the geometry is Ricci flat.  All the information is in the Gibbons-Hawking term evaluated on the boundary cut-off surface. 
One finds
\begin{equation}\label{eq:kerrKgam}
\begin{split}
K &= \frac{1}{2\,\sqrt{\Delta}\,\rho^3}\,\dv{r}(\rho^2\,\Delta) = \frac{2}{r} - \frac{M}{r^2} + \cdots\,,\\
\sqrt{\gamma}&= \sin\theta\, \sqrt{\Delta}\, \rho \,.
\end{split}
\end{equation}
Rather than include counterterms in the asymptotically flat region, we perform a background subtraction, and evaluate 
\begin{equation}
I_{_\text{EH,on-shell}} = -\frac{2}{16\pi\,G_N}\int_0^\beta\,\d t\,\int_0^\pi\,\d \theta\, \int_0^{2\pi}\,\d \phi \, \sqrt{\gamma}\left[K - K_0\right]
\end{equation}	
where $K_0$ is the extrinsic curvature trace for the solution with $M=0$ (flat space). This procedure leads to a finite result for the grand canonical free energy $G$
\begin{equation}\label{eq:kerrSos}
\mathcal{Z}[\beta,\Omega] = e^{-\beta\,G }= e^{-I_{_\text{EH,on-shell}}} \,, 
\qquad  \quad 
I_{_\text{EH,on-shell}}  = \frac{M\,\beta}{2\,G_N}\,.
\end{equation}	
The canonical free energy $F$ in turn is obtained by a Legendre transform 
\begin{equation}\label{eq:KerrF}
F = G - \Omega \,J = \frac{r_+-M}{2\,G_N}\,.
\end{equation}	
As a sanity check note that as $a\to0$, $r_+ \to 2\,M$ and we obtain correctly the Schwarzschild free energy. Since this calculation is valid for a generic Kerr black hole, we can easily evaluate the near-extremal result simply by expanding~\eqref{eq:KerrF} for small $T$ using our parameterization in~\eqref{eq:MalowT} to find
\begin{equation}
F = \frac{\Js^2\,\Ts}{2\,G_N}(1+\Js\, \Ts + \cdots)\,.
\end{equation}	
We note in passing that the grand canonical free energy also has a sensible low $T$ expansion, since $G \propto M$. 

To evaluate the on-shell action of the near-extremal solution, we work with the solution parameterized as in~\eqref{eq:dsq01alt}. The extremal near-horizon has \AdS{2} asymptotics. However, the leading temperature correction is an irrelevant deformation of these asymptotics, as is clear from the line element $\d s_1^2$. Therefore, the evaluation will require careful regulating of the divergences. 

Let us consider the canonical ensemble with fixed asymptotic charge. We have to implement the Legendre transformation as in~\eqref{eq:KerrF} to fix the charge, as opposed to the holonomy of the gauge field. This is easy to do: we include a counterterm that changes the boundary condition for the gauge field from Dirichlet (fixed holonomy) to Neumann (fixed charge). Naively, we would include this counterterm in the asymptotically flat region, but in the case of the fixed charge ensemble, the radial Gauss law implies that the selfsame counterterm suffices in the near-horizon region. In fact, this Neumann boundary condition is natural from the near-horizon \AdS{2} perspective, since the gauge potential grows linearly, $A_t \sim y$, as $y\to \infty$.\footnote{ As discussed in~\cite{Iliesiu:2022onk} one can work in the grand canonical ensemble by imposing a mixed boundary condition.} In short, once we zoom into the near-horizon region, in the canonical ensemble, we fix the metric and the field strength at a fixed constant $y$ hypersurface.

While this Neumann boundary term suffices at strict extremality to regulate the on-shell action, one must further regulate the divergence arising from the 
irrelevant deformation of \AdS{2} asymptotics. Since we are holding the metric fixed, we have the freedom to include counterterms built from the induced first metric on the fixed $y$ hypersurface. Following usual AdS counterterm analysis we can see that the natural counterterms are a hypersurface cosmological constant, and a hypersurface Einstein-Hilbert term. Putting all this together, we have
\begin{equation}
I_{_\text{EH}}^\text{reg} = I_{_\text{EH}} - \frac{1}{16\pi G_N}  \int\, \dd^3x\, \sqrt{\gamma}\, \left(\alpha\, n_\mu \,A_\nu\, F^{\mu\nu}  + c_1 +  c_2 \, {}^\gamma\!R \right)  
\end{equation}	

The coefficient of the Neumann boundary term $\alpha$ is fixed by demanding that the zero temperature calculation in the near-horizon region lead to a finite result. This determines $\alpha = \frac{8}{\pi}$. It also imposes a linear relation between $c_1$ and $c_2$, since each of these terms gives a linearly divergent action. Including the leading temperature correction, we find a quadratically divergent result from $I_{_\text{EH}}$.  At this stage we 
exploit the freedom to the constant mode of the gauge potential $A_\mu$ at $\order{\Ts}$ (which does not affect the field strength), and tune the relation between $c_1$ and $c_2$. The specific values of the counterterms and the shift of the background gauge potential are not particularly illuminating, and so we refrain from quoting them.   The upshot of this analysis is the result quoted in~\eqref{eq:SosT}.  

%~~~~~~~~~~~~~~~~~~~~~~~~~~~~~~~~~~~~~~~~ 
\section{Compensating diffeomorphism for vector mode} \label{sec:rotApp}
%~~~~~~~~~~~~~~~~~~~~~~~~~~~~~~~~~~~~~~~~ 

According to~\cite{Sen:2012cj}, the zero modes of the extremal Kerr black hole geometry are associated to large diffeomorphisms that act on the boundary of the throat. Their concrete profile in the bulk is sensitive to gauged diffeomorphisms that fall off fast enough near the boundary. Therefore, the precise profile of the zero modes is uniquely determined only after a choice of gauge fixing term in the action. The naive zero modes are associated to the isometries of the \AdS{2} factor in the geometry, namely the tensor modes analyzed in~\cref{sec:tensorz}, and to the rotation around $\varphi$. The purpose of this Appendix is to show that, working in harmonic gauge as in Eq.~\eqref{eq:gaugefix}, the rotational (would-be) zero mode has actually a non-zero action even at zero temperature. This is the puzzle explained in~\cref{sec:rotatez}.

We shall begin with some general comments that will simplify the analysis below. The Einstein-Hilbert action including the Gibbons-Hawking-York boundary term can be expanded to quadratic order around a solution of the equations of motion. We denote the kernel by $\widetilde{\Delta}_L$, such that 
\begin{equation}
    -\int \d^4 x \sqrt{g} \,R - 2 \oint  \d^3 x \sqrt{\gamma} \,K = I_{\rm classical} +\int \d^4x\, \sqrt{g} \, \frac{1}{4}\,  \widetilde{h}^{\mu\nu} \widetilde{\Delta}_L\, h_{\mu\nu} + \mathcal{O}(h^3)
\end{equation}
The relation between this operator and the one defined through~\eqref{eq:DeltaLgen} is
\begin{equation}
 \Delta_L h_{\mu\nu} = \widetilde{\Delta}_L h_{\mu\nu}- \underbrace{2  \nabla_\mu \nabla_\rho h^{\rho}{}_{\nu} + \nabla_\mu \nabla_\nu h^{\rho}{}_\rho + g_{\mu\nu} \left( \nabla_\rho \nabla_\sigma h_{\rho \sigma} - \frac{1}{2} \nabla_\rho \nabla^\rho h^\sigma{}_\sigma \right)}_{\rm From ~ Gauge-Fixing~Term}.
\end{equation}
One can check explicitly that $\widetilde{\Delta}_L \nabla_\mu \xi_\nu = 0$ as expected from a diffeomorphism invariant action. 
The rest of the right hand side comes purely from the gauge fixing term $I_\text{gf} = \frac{1}{32\pi G_N}\, \int \d^4x\, \sqrt{g} \,{\sf GF}_\mu {\sf GF}^{\mu},$ where for convenience we defined 
\begin{equation}
    {\sf GF}_\mu \equiv \nabla_\rho h^{\rho}{}_\mu- \frac{1}{2} \nabla_\mu h^\rho{}_\rho.
\end{equation}

Instead of finding the zero modes of $\Delta_L$, in this section we will look for diffeomorphism $\xi$ such that ${\sf GF}_\mu = 0$, which automatically implies that
\begin{equation}
    {\sf GF}_\mu = 0  \;\; \Longrightarrow \;\; \Delta_L \nabla_\mu \xi_\nu =0.
\end{equation}
After these preliminaries we can start our analysis. First, we shall remind the reader that the naive rotational vector zero mode is generated by a linear combination of diffeomorphisms
\begin{equation}\label{eq:vectortry1}
\textit{Ansatz 1:}\qquad \xi_R^\mu = e^{\i n \tau} f(y) \, (\partial_\varphi)^\mu \,.
\end{equation}
In analogy with the Schwarzian modes we expect $f(y) \propto (\frac{y-1}{y+1})^{\frac{n}{2}}$. Our main observation is that this transformation is not in the harmonic gauge since 
\begin{equation}
{\sf GF}_y = e^{\i n \tau} \frac{4 n \sin^2 \theta}{(y^2-1)(1+\cos^2 \theta)^2} f(y),\quad\Rightarrow \quad f(y)=0.
\end{equation}
This implies that there is no diffeomorphism taking the form~\eqref{eq:vectortry1} that is a zero mode of $\Delta_Lh_{\mu\nu}$. The second option we could try is 
\begin{equation}\label{eq:casealsointensor}
\textit{Ansatz 2:} \qquad \xi_R^\mu =e^{\i n \tau} \left(f_1(y)\, \pdv{y} + f_2(y) \pdv{t} + f_3(y) \pdv{\varphi} \right).
\end{equation}
Nevertheless, we already proved in~\cref{sec:tensorz} that the only diffeomorphism of this form that generates a zero mode of $\Delta_L h_{\mu\nu}$ are the Schwarzian modes themselves and not the new rotational zero mode we are looking for. Next, we consider a generalization
\begin{equation}
\textit{Ansatz 3:}\qquad \xi_R^\mu =e^{\i n \tau} f_\mu(y)\, \pdv{x^\mu}.
\end{equation}
This adds a $(y,\tau)$-dependent generator along $\theta$ compared to the previous case. When evaluating the gauge fixing term we find $(\sin^2\theta \,{\sf GF}_\theta )|_{\theta\to 0 } = - e^{\i n \tau} f_\theta(y)$ and therefore $f_\theta(y)=0$, getting us back to the previous case~\eqref{eq:casealsointensor}.

Since we find no $(y,\tau)$-dependent diffeomorphism that produces a zero mode, we next allow for $\theta$-dependence along some coordinates. We assume adding a dependence with $\varphi$ will never result in a zero mode. A reasonable generalization of our ansatz since we know the wave equation around the Kerr geometry to be separable is therefore 
\begin{equation}
\textit{Ansatz 4:} \qquad 
\xi_R^\mu  =e^{\i n \tau} \left(\alpha(\theta)f_1(y)\, \pdv{y} + \alpha(\theta) f_2(y) \pdv{t} + f_3(y) \pdv{\varphi} \right).
\end{equation}
One can show once again no such diffeomorphism exists such that ${\sf GF}=0$. The procedure is a bit more involved, so we will only outline it. First, we look at ${\sf GF}_\theta=0$ which determines $f_2(y)$ in terms of $f_1(y)$. Next we evaluate ${\sf GF}_y$ in an expansion around $\theta = 0$. We find that in order for ${\sf GF}_y = 0$ order by order in $\theta$ we need to impose $\alpha'(0) = 0$, $\alpha''(0) = \alpha(0)/2$, and $\alpha'''(0)=\alpha'''''(0) = 0$ but the equations that determine $\alpha^{(4)}(0)$ and $\alpha^{(6)}(0)$ are incompatible leaving $\alpha^{(n)}(0)=0$ as the only solution for $n=0,\ldots,6$. Carrying this out to higher orders we expect the only solution to be $\alpha(\theta)=0$. At that point we are back to one of the previous cases.

Instead of following this route to try and generalize the angle dependence, we found the following choice to simplify the result. Take the diffeomorphism to be 
\begin{equation}
\textit{Ansatz 5:} \qquad 
\xi_R^\mu = e^{\i n \tau} f_4(y) \, \Big(\pdv{\varphi}\Big)^\mu  + \xi_H^\mu \,, 
\quad \xi^H_\mu\, \d x^\mu = e^{\i n \tau} (f_1(y) \d y + f_2(y) \d \tau + f_3(y) \d \varphi).
\end{equation}
This combination of the naive rotational zero mode proportional to $\partial_\varphi$ and a compensating diffeomorphism. The latter is presented  in terms of its dual one-form to simplify the evaluation of ${\sf GF}_\mu$. When the components are written down explicitly, this choice introduces dependence on $\theta$ through the metric factor needed to raise the index of $\xi_H$. After some tedious calculation we can show that the only solution is $f_1(y) = f_2 (y) = f_3 (y) = f_4 (y) = 0$. To see this first determine $f_2(y)$ from ${\sf GF}_\theta=0$. Then $f_3(y)$ can be determined from ${\sf GF}_\varphi(\theta=0)=0$. From ${\sf GF}_y (\theta=0)=0$ and smoothness at the horizon we can find $f_1(y)$ and finally $f_4(y)$ from ${\sf GF}_y (\theta=\frac{\pi}{2})=0$. After this we can verify the solution is not in the harmonic gauge unless all $f$'s vanish. 

Finally, we try the last ansatz, using which we do find a solution. This is the one quoted in~\cref{sec:rotatez}, which happens to be singular. 
\begin{equation}
\textit{Ansatz 6:}\qquad \xi_R^\mu = e^{\i n \tau} f_4(y) \pdv{\varphi}  + \xi_H^\mu,
\qquad \xi^H_\mu \,\d x^\mu = e^{\i n \tau} \alpha(\theta)(f_1(y) \d y + f_2(y) \d \tau ).
\end{equation}
This solution can be ruled out using similar tricks as in the previous cases. Nevertheless, this gets the closest to a possible solution, provided we are willing to live with a singularity on the axis. For example,  using the ansatz 
\begin{equation}\label{eq:A6sol}
f_4(y) = c \left(\frac{y-1}{y+1}\right)^{\frac{n}{2}}\,, \quad
f_1(y) = \partial_y\, f(y) \,, \quad f_2(y) = \i \, n \, f(y) \,,
\quad f(y) \equiv  \left(\frac{y-1}{y+1}\right)^{\frac{n}{2}}\,,
\end{equation}
where $c$ is an arbitrary coefficient, we can put this diffeomorphism in the harmonic gauge as long as 
\begin{equation}
\alpha(\theta) =  c_1 + c_2 \left( \cos\theta+2\,\log\tan\frac{\theta}{2} \right) + 4\, \log \sin\theta\,.
\end{equation}
This is the result quoted in~\cref{sec:rotatez}. There is no choice of $c_1$ and $c_2$ such that $\alpha(\theta)$ is smooth along the sphere. We tried adding a component along $\theta$ to the compensating diffeomorphism $\xi^H_\mu\, \d x^\mu \to \xi^H_\mu\, \d x^\mu+ e^{\i n \tau} \,\beta(\theta)\,  (\frac{y-1}{y+1})^{\frac{n}{2}}\, \d \theta$ but found $\beta(\theta)=0$ to be the only smooth solution regardless of $\alpha(\theta)$.

Incidentally, if we only seek a compensating vector field $\xi_H$ that brings our guess $\xi_R^\mu = \left(\frac{y-1}{y+1}\right)^{\frac{n}{2}}\, e^{\i n\tau}\,\pdv{\varphi}$ to harmonic gauge, we can do so in a slightly simpler manner. Demanding that ${\sf GF}_\mu =0$ for a diffeomorphism along $\xi_R + \xi_H$, implies that $\xi_H$ satisfy
\begin{equation}
    \nabla_\mu\nabla^\mu \xi^H_\rho = \frac{4\,g_2}{g_1} \left(\partial_y f \,\d y+ i\,n\, f\, \d \tau\right),
\end{equation}
with $f(y)$ given in~\eqref{eq:A6sol}. This equation can be solved directly to recover the vector field quoted in the main text in~\cref{sec:rotatez}. This is, of course, no different from what was described above, but it is helpful to note that the compensating diffeomorphism satisfies an inhomogeneous Laplace equation.

To summarize the discussion, we were unable to find a compensating diffeomorphism that puts the naive rotational zero mode, that generates \AdS{2} dependent translations along $\varphi$, in harmonic gauge. Nevertheless, we should stress we cannot rule out the possibility of a complicated solution which is not part of the ansatz explored above, although we find it unlikely. 

%%%%%%%%%%%%%%%%%%%%%%%%%%%%%%%%%%%%%%%%%%%%%%%%%%%%%%  
\bibliographystyle{JHEP}
\bibliography{kerr-refs}
\end{document}